\documentclass[twocolumn,secnumarabic,amssymb, nobibnotes, aps, pra]{revtex4-1}

\newcommand{\os}{\ensuremath{\omega_\mathrm{s}}}
\newcommand{\oi}{\ensuremath{\omega_\mathrm{i}}}
\newcommand{\oin}{\ensuremath{\omega_\mathrm{in}}}
\newcommand{\oout}{\ensuremath{\omega_\mathrm{out}}}

\usepackage{braket}
\usepackage{graphicx}
\usepackage{color,soul}
\usepackage{amsmath}
\usepackage{amssymb}

\begin{document}

\title{Tailoring nonlinear processes for quantum optics \\ with pulsed temporal-mode encodings}

\author{Vahid Ansari$^{1}$}
\email{vahid.ansari@uni-paderborn.de}
\author{John M. Donohue$^{1}$}
\author{Benjamin Brecht$^{2}$}
\author{Christine Silberhorn$^{1}$}

\affiliation{$^1$Integrated Quantum Optics, Paderborn University, Warburger Strasse 100, 33098 Paderborn, Germany}
\affiliation{$^2$Clarendon Laboratory, University of Oxford, Parks Road, Oxford, OX1 3PU, UK}

\begin{abstract}
The time-frequency degree of freedom is a powerful resource for implementing high-dimensional quantum information processing. In particular, field-orthogonal pulsed temporal modes offer a flexible framework compatible with both long-distance fibre networks and integrated waveguide devices. In order for this architecture to be fully utilised, techniques to reliably generate diverse quantum states of light and accurately measure complex temporal waveforms must be developed. To this end, nonlinear processes mediated by spectrally shaped pump pulses in group-velocity engineered waveguides and crystals provide a capable toolbox. In this review, we examine how tailoring the phasematching conditions of parametric downconversion and sum-frequency generation allows for highly pure single-photon generation, flexible temporal-mode entanglement, and accurate measurement of time-frequency photon states. We provide an overview of experimental progress towards these goals, and summarise challenges that remain in the field.
\end{abstract}

\maketitle

\section{Introduction}\label{sec:into}

In any implementation of quantum information protocols, it is necessary to have access to information-carrying modes that are individually manageable and measurable in arbitrary bases. In optical implementations, it is often essential to be able to create photonic quantum states with a controlled degree of entanglement and to retain coherence among the modes over long-distance transmission. In polarisation, state rotations and measurements are simple with wave-plates and polarising beam splitters, and entangled sources are straightforward to implement, but the dimensionality is limited to two. In the spatial degree of freedom, entanglement is naturally present in a high-dimensional basis of, for example, orbital angular momentum modes, and arbitrary measurements can be made with spatial light modulators. However, their complex spatial structures render them incompatible with spatially single-mode integrated devices and optical fibre networks.

\begin{figure}[!b]
\centering
\includegraphics[width=1\linewidth]{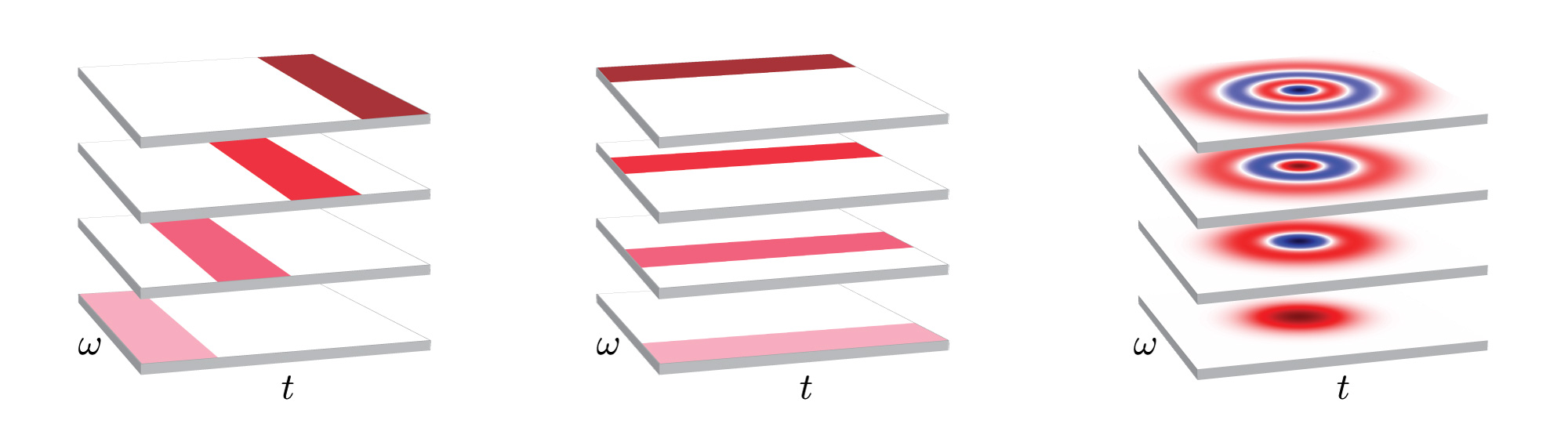}
\caption{Temporal-mode encodings visualised in time-frequency space. Orthogonal temporal mode bases can be constructed through slicing bins in time or frequency, as in (a) and (b), or through intensity-overlapping but field-orthogonal pulsed temporal modes, such as the Hermite-Gauss modes in (c).}
\label{fig:wigner}
\end{figure}

Alternatively, the time-frequency (or energy-time) degree of freedom can be exploited by encoding quantum information in photonic \textit{temporal modes} (TMs). Here, the information is encoded in the complex time-frequency amplitude of the electric field of single photons. Like spatial encodings, the Hilbert space available in the Fourier-conjugate time and frequency domains is in-principle unbounded, allowing for high-dimensional encodings. Unlike spatial encodings, time-frequency encodings are intrinsically compatible with waveguides and fibre transmission. Temporal-mode bases can take on a variety of forms, such as discrete time or frequency bins or intensity-overlapping pulsed temporal modes, as illustrated in Fig \ref{fig:wigner}, so long as the waveforms provide an orthonormal basis. However, controlling entanglement between and directly measuring arbitrary temporal modes presents a significant challenge for time-frequency quantum information processing.

\begin{figure*}[ht]
\centering
\includegraphics[width=1\textwidth]{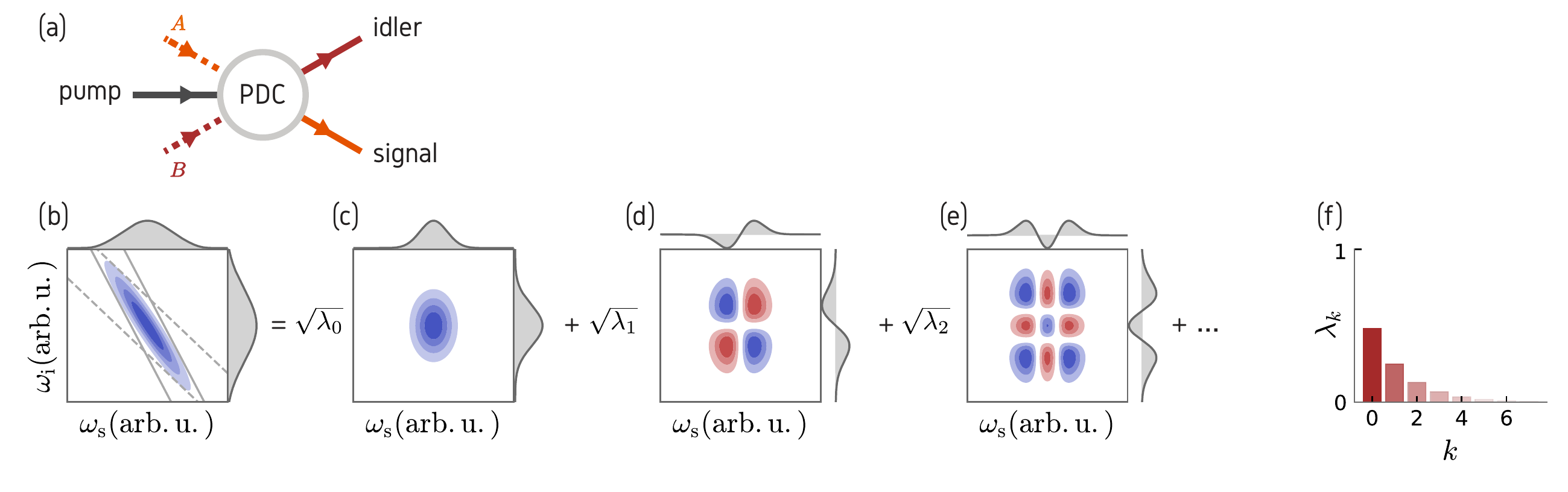}
\caption{Joint spectral amplitude, temporal modes, and Schmidt coefficients of a non-engineered PDC process. (a) Outline of a PDC process with the three involved fields. (b) The JSA and its marginal distributions which is the product of pump (dashed lines) and phasematching (solid lines) functions and, in this case, exhibits frequency anti-correlations between signal and idler frequencies. The Schmidt decomposition of this Gaussian JSA is given by Hermite-Gaussian functions, with the first three TM pairs shown in (c-e). (f) The first seven Schmidt coefficients $\lambda_k$. The decomposition of this example yields an effective mode number of $K\approx3.14$.}
\label{fig:jsa}
\end{figure*}

In this review, we will highlight works on both the targeted generation and manipulation of TMs through controlling the group-velocity relationship in nonlinear processes. In Section~\ref{sec:one}, we summarise the basic theory behind the TM structure of photon pairs generated via parametric downconversion (PDC). Section~\ref{sec:two} focusses on efforts towards engineering the PDC process itself, for both single-mode photon generation and to create photons with rich, programmable TM structures. In Section~\ref{sec:three}, we transfer these techniques from PDC to frequency conversion, unveiling methods to manipulate and measure the complex TM structure. Section~\ref{sec:four} then summarises current experimental progress on the manipulation of photonic TMs by means of frequency conversion, direct temporal manipulation, and tailored light-matter interactions. In Section~\ref{sec:five}, we overview recent experimental results paving the way towards TM-based quantum applications. Finally, in Section~\ref{sec:conc}, we will give an outlook on future steps and highlight challenges that will need to be overcome in the future.

\section{Temporal-mode structure of parametric down-conversion}\label{sec:one}
In this section we describe the TM structure of photon-pair states generated in PDC, where a photon from a bright classical pump pulse decays with a small probability inside a nonlinear optical medium, e.g. a nonlinear waveguide, into a pair of daughter photons typically called \textit{signal} and \textit{idler}, as sketched in Fig.~\ref{fig:jsa}a. PDC is a well-understood process, capable of generating photons with a rich TM structure at room temperature. Moreover, PDC can be used to generate a plethora of quantum states, including heralded single photons, squeezed states, and maximally entangled states. These properties have cemented PDC as the workhorse in many quantum optics laboratories.

Restricting our model to the generation of photon pairs and assuming spatially single-mode emission, e.g. by realising the PDC in a weakly pumped waveguide, the type-II PDC process can be described by the interaction Hamiltonian
\begin{equation}
\hat{H}_{\mathrm{PDC}}= \mathcal{B} \int d\os\,d\oi\,f(\os,\oi)\hat{a}^\dagger(\os)\hat{b}^\dagger(\oi) + \mathrm{h.c.},
\end{equation}
and the generated state can be written as
\begin{equation}
|\psi\rangle_\mathrm{PDC}= \mathcal{B} \int d\os\,d\oi\,f(\os,\oi)\hat{a}^\dagger(\os)\hat{b}^\dagger(\oi)|vac\rangle,
\end{equation}
where $\hat{a}^\dagger(\os)$ and $\hat{b}^\dagger(\oi)$ are standard creation operators that generate a signal photon at $\os$ and an idler photon at $\oi$, $\mathcal{B}$ is the optical gain or efficiency of the process which includes the second-order nonlinearity and the pump power, and $f(\os,\oi)$ is the complex-valued joint spectral amplitude (JSA), normalised to ${\int d\os\,d\oi\,|f(\os,\oi)|^2=1}$. The JSA describes the entangled time-frequency structure of the PDC state, and is essential for describing PDC in cases with a broadband pump pulse~\cite{Grice1997}.

The JSA itself can be written as a product of the pump envelope function $\alpha(\os+\oi)$ and the phasematching function $\phi(\os,\oi)$, such that
\begin{equation}
f(\os,\oi) = \alpha(\os+\oi) \phi(\os,\oi).
\label{eq:jsa}
\end{equation}
Here, $\alpha(\os+\oi)$ is the slowly varying envelope of the broadband pump and reflects energy conservation during the PDC, and the phasematching $\phi(\os,\oi)$ expresses the momentum conservation between involved fields and the dispersion properties of the nonlinear medium. The phasematching function can be written as
\begin{equation}
\phi(\os, \oi) = \int_0^L dz\,\chi(z)\exp\left[\imath\Delta k(\os,\oi)z\right],
\label{eq:phasematching0}
\end{equation}
where $\Delta k(\os,\oi) = k_\mathrm{p}(\os + \oi) - k_\mathrm{s}(\os)-k_\mathrm{i}(\oi)$ is the phase mismatch, $L$ is the length of the nonlinear medium, and ${\chi(z)=\pm1}$ describes the orientation of the ferroelectric domains of the crystal.  A periodic modulation of $\chi(z)$, with a period $\Lambda$, is called \textit{periodic poling}~\cite{hum2007quasi}. This poling adds an additional component of the form ${k_\mathrm{QPM}=2\pi/\Lambda}$ to the phase mismatch such that ${\Delta k(\os,\oi) \mapsto \Delta k(\os,\oi) + 2\pi/\Lambda}$, allowing the centre frequencies of the phasematched process to be tuned. In this case, the resulting phasematching function is given by
\begin{equation}
\phi(\os, \oi)=\frac{1}{L}\mathrm{sinc}\left(\frac{\Delta k(\os, \oi)L}{2}\right) e^{\imath\Delta k(\os, \oi)\frac{L}{2}}.
\label{eq:phasematching1}
\end{equation}
The sinc profile of the phasematching function has significant implications which will be discussed in Section~\ref{sec:two}. However, to simplify the equations and plots in this article, we usually employ a Gaussian approximation of the phasematching function.

In 2000, Law and co-workers examined the time-frequency structure of the JSA through the \textit{Schmidt decomposition}, defining two-photon entanglement in terms of temporal modes~\cite{Law2000}. For this, the JSA is decomposed into two sets of orthonormal basis functions $\{g^{(s)}\}$ and $\{h^{(i)}\}$ for signal and idler, and we write
\begin{equation}
f(\os,\oi)=\sum_k\sqrt{\lambda_k}g_k^{(s)}(\os)h_k^{(i)}(\oi),
\end{equation}
where $\sum_k\lambda_k=1$. With this we define broadband TM operators 
\begin{align}
\hat{A}^\dagger_k &= \int d\os\,g_k^{(s)}(\os)\hat{a}^\dagger(\os),\\
\hat{B}^\dagger_k &= \int d\oi\,h_k^{(i)}(\oi)\hat{b}^\dagger(\oi),
\end{align}
and consequently obtain 
\begin{equation}
|\psi\rangle_\mathrm{PDC} = \sum_k\sqrt{\lambda_k}\hat{A}_k^\dagger\hat{B}_k^\dagger|0\rangle,
\label{eq:pdc_schmidt}
\end{equation}
where we have postselected on and renormalised for two-photon emission. This means that given a PDC photon pair is generated, it is in the $k$-th TM pair with a probability of $\lambda_k$. An example of a typical JSA together with its Schmidt decomposition is given in Fig. \ref{fig:jsa}b. For a typical Gaussian JSA, the Schmidt modes are given by Hermite-Gauss functions, which overlap in both spectral and temporal intensity.

The Schmidt decomposition of the joint spectral amplitude provides an essential link between the continuous time-frequency description and a discretised temporal-mode picture. Such a transition is necessary for describing mode-multiplexed systems, where each Schmidt mode can be thought of as an independent information carrier. Such multiplexed systems are useful for communication networks~\cite{ghalbouni2013experimental} and essential to generate highly entangled cluster states for measurement-based quantum computation~\cite{menicucci2006universal,pysher2011parallel,yokoyama2013ultra,chen2014experimental}, where utilising the time-frequency domain allows for operations to take place in a single spatial mode. The Schmidt modes of PDC can be directly connected to the supermodes generated in a synchronously pumped optical parametric oscillator (SPOPO), where a degenerate downconversion medium is pumped below threshold in a cavity matched to the repetition rate of the driving laser system~\cite{DeValcarcel2006,Pinel2012}. The eigenmode decomposition of the interaction provides the independently squeezed supermodes of the system~\cite{wasilewski2006pulsed,Patera2012}, and their mixtures have been experimentally demonstrated to exhibit strong continuous-variable entanglement~\cite{Roslund2014,Gerke2015}.

In the low-gain PDC regime, the Schmidt decomposition of the JSA can be linked directly to the amount of time-frequency entanglement present in the two-photon system. The Schmidt number, defined as $K=1/\sum_k\lambda_k^2$, quantifies the number of TM pairs required to describe the properties of the generated state, with $K=1$ for a single-mode (separable) state and $K\gg1$ for a multimode (entangled) state~\cite{huang1993correlations,Uren2005,mikhailova2008biphoton}. The Schmidt number is related to the spectral purity of the individual signal photons generated, which are generally described by the mixed density matrix \begin{equation}\hat{\rho}_\mathrm{s}=\mathrm{Tr}_\mathrm{i}(\hat\rho_\mathrm{PDC})=\sum_k\lambda_k|A_k\rangle\langle A_k|\end{equation} with a purity of \begin{equation}\mathcal{P}_\mathrm{s}= tr(\hat{\rho}_\mathrm{s}^2) = \frac{1}{K}.\label{eq:purity}\end{equation} For PDC-generated photons, this quantity is directly experimentally accessible through the marginal second-order correlation function (i.e. unheralded signal photons) as ${g^{(2)}(0)=1+\mathcal{P}_\mathrm{s}}$~\cite{Christ2011,Eckstein2011,wakui2014ultrabroadband}.

In summary, we have introduced the continuous time-frequency structure of PDC and connected it to the discrete TM picture through the Schmidt decomposition. Such analysis naturally describes the two-photon entanglement from PDC, the squeezed modes of a pulsed OPO, and the spectral purity of the generated photons. In most configurations, PDC generates highly correlated states with a large Schmidt number, yielding low-purity heralded photons if no additional spectral filtering is applied. We will shift our focus in the next section to how proper engineering of the PDC process can overcome this limitation and facilitate the direct generation of pure single photons.

\section{PDC engineering}\label{sec:two}
Although multimode PDC states with usual frequency anti-correlations, as shown in Fig.~\ref{fig:jsa}, have found many applications in quantum science~\cite{nasr2003demonstration,giovannetti2001quantum,averchenko2017temporal}, full control over the modal structure of the PDC state would make a new range of applications possible. For example, high-visibility quantum interference between distinct nodes in a photonic network requires pure PDC sources, i.e. sources that emit in a single temporal mode. Without dispersion engineering, intrinsic frequency anti-correlations between signal and idler are imposed by energy conservation of the pump, reflected by the $-45^\circ$ angle of the pump function in the joint spectral amplitude (see Fig.~\ref{fig:jsa}), resulting in highly multimode systems. To realise single-mode PDC, researchers have tailored the phasematching function to produce separable JSAs, allowing for high-quality heralded photons without any need for additional spectral filtering.

\begin{figure}
\centering
\includegraphics[width=.9\linewidth]{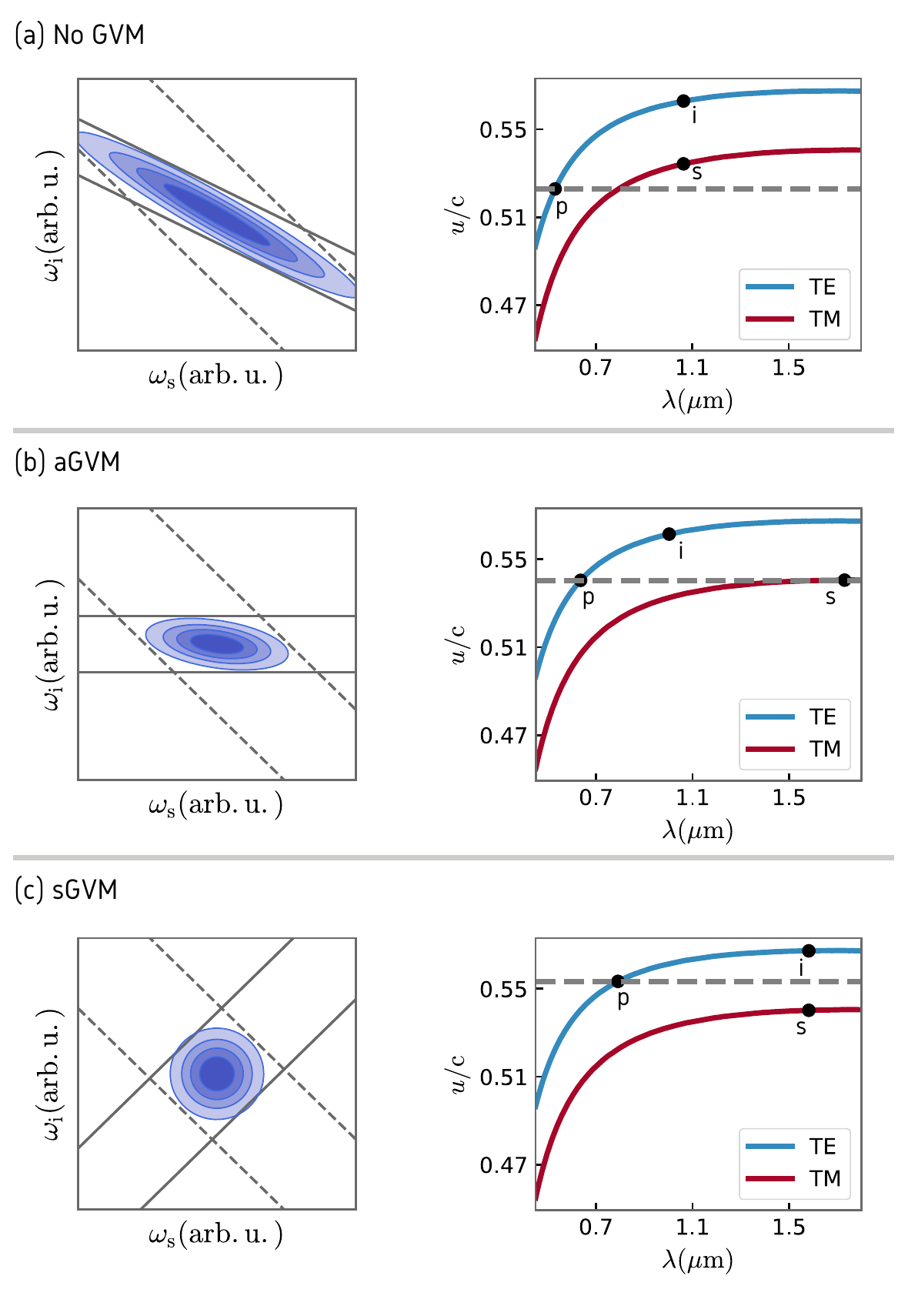}
\caption{Three different group-velocity matching condition. The JSA of each case is plotted on the left side, with the respective group velocities $u_j$ of the pump, signal, and idler fields plotted on the right side. The group velocities (normalised over the speed of light in vacuum) are exemplary for TE and TM-polarised light in a z-cut KTP crystal. (a) Typically without dispersion engineering, the long-wavelength signal and idler photons both have a larger group velocity than the pump ($\xi>0$). This leads to a negative phasematching angle and consequently to a correlated JSA as shown on the left. In this example, $\xi\approx0.4$. (b) In the case of aGVM ($\xi\rightarrow0$), one photon (here the signal) propagates at the same velocity as the pump. This yields a phasematching function that is aligned with the signal or idler frequency axis. If the pump spectral bandwidth is larger than the phasematching bandwidth, a separable JSA is generated. (c) For sGVM ($\xi\rightarrow-1$), the group velocity of the pump lies between the group velocities of signal and idler. This leads to a $+45^\circ$ phasematching angle and, given that the pump spectral bandwidth matches the phasematching bandwidth, a separable JSA with potentially indistinguishable signal and idler.}
\label{fig:gv}
\end{figure}

\subsection{Group-velocity matching for single-mode emission}
At the turn of the millennium, several groups studied the spectral characteristics of PDC photon pairs and identified a connection between the photon spectra and the dispersion of the nonlinear medium~\cite{Keller1997,Erdmann2000,Giovannetti2002}. It was shown that  with a properly selected non-linear material, polarisations, and photon central frequencies the frequency correlations between the signal and idler photons can be eliminated~\cite{Grice2001}. Later this work was further developed in~\cite{Uren2005}, where the authors showed that the relationship between the group velocities of interacting fields plays an essential role in tailoring the phasematching function $\phi(\os, \oi)$ and consequently the JSA. 

To understand the underlying physics, we perform a Taylor expansion of the phase mismatch (defined in Sec.~\ref{sec:one}) up to the first order. Assuming that the process is perfectly phasematched at the centre frequencies and that group-velocity dispersion through the nonlinear medium is negligible, we obtain 
\begin{equation} 
\Delta k(\os,\oi) \approx (u^{-1}_{\mathrm{s}}-u^{-1}_{\mathrm{p}})\os + (u^{-1}_{\mathrm{i}}-u^{-1}_{\mathrm{p}}) \oi,\label{eq:delKpdc1stord} 
\end{equation}
where the $u_j \equiv \frac{\partial\omega_j}{\partial k_j}$ are the group-velocities of the pump, signal, and idler fields. In this context, it is useful to define the group-velocity-mismatch contrast $\xi$ as
\begin{equation}
\xi = \frac{ u^{-1}_{\mathrm{s}} - u^{-1}_{\mathrm{p}} }{ u^{-1}_{\mathrm{i}} - u^{-1}_{\mathrm{p}} }.
\label{eq:gvmc}
\end{equation}
The group-velocity mismatch contrast is related to the angle of the phasematching function in the $(\os,\oi)$-plane by $\theta_\mathrm{PM}=-\arctan(\xi)$~\cite{Uren2005}.

Among all possible group-velocity arrangements, two special cases received particular attention. In the first case, dubbed \textit{asymmetric group-velocity matching} (aGVM), the pump propagates with the same group velocity as either the signal photon ($\xi\to0$) or the idler photon ($\xi\to\infty$). If the pump is group-velocity matched to the signal photon, the JSA from Eq. (\ref{eq:jsa}) is reduced to
\begin{equation}f(\os,\oi) \approx \alpha(\os+\oi) \phi(\oi).\label{eq:jsa_aGVM}\end{equation} As seen in Fig. \ref{fig:gv}(b), as the phasematching bandwidth shrinks to be much narrower than the pump bandwidth, the JSA becomes more and more separable. The single-modedness of the system can be increased by using wider pump bandwidths or tightening the phasematching function with longer nonlinear interactions~\cite{Uren2005}. In this scenario, the signal and idler photon will have drastically different spectral bandwidths.

In the second case, the group velocity of the pump is exactly between the group velocities of signal and idler ($\xi\to-1$), referred to as \textit{symmetric group-velocity matching} (sGVM) or \textit{extended phasematching}, which results in a JSA of the form \begin{equation}f(\os,\oi) \approx \alpha(\os+\oi) \phi(\os-\oi).\label{eq:jsa_sGVM}\end{equation} As seen in Fig.~\ref{fig:gv}(c), if the phasematching bandwidth equals the pump bandwidth, the JSA is a perfectly separable circle, allowing for pure single photons with identical spectral properties. This phasematching configuration also allows for two-photon states with positive spectral correlations (and negative temporal correlations) when the pump is broader than the phasematching function~\cite{Kuzucu2005,lutz2014demonstration,Ansari2014}, useful for certain quantum synchronisation and dispersion-cancellation techniques.

\subsection{Experimental high-purity photon sources}

The first experimental demonstrations of separable photon-pair generation were realised in nonlinear bulk crystals. In these systems, the spatial and spectral properties of the photon pairs can be linked during generation, depending on the focus of the pump and collection optics~\cite{bennink2010optimal}. In 2007, the group of J. P. Torres demonstrated control over the spectral correlations using this spectral-spatial coupling for photon pairs at 810~nm generated in LiIO$_3$~\cite{Valencia2007}. By adapting the spatial mode of the pump, the generated photon pairs could be tuned from spectrally correlated to separable. This was verified by a measurement of the joint spectral intensity (JSI), $|f(\os,\oi)|^2$.

In 2008, the group of I. A. Walmsley demonstrated engineered PDC under aGVM conditions in a bulk KDP crystal~\cite{Mosley2008}, with photon pairs produced around 830~nm. In addition to JSI measurements, the authors demonstrated Hong-Ou-Mandel interference~\cite{Hong1987} between heralded photons from two different PDC sources with a visibility of 94.4\%.

Also in 2008, F. Wong's group designed and analysed a source of telecom photon pairs produced under sGVM conditions in periodically poled KTiOPO$_4$ (ppKTP) crystal~\cite{Kuzucu2008}. To measure correlations, the photons were up-converted in a second nonlinear crystal by a short gate pulse. By scanning the relative delay of the photons and the gate, the authors were able to measure the joint \textit{temporal} intensity, explicitly showing the possibility of temporal anti-correlation under sGVM conditions. This demonstrated for the first time that changing the spectral bandwidth of the pump facilitates control over the time-frequency correlations of the pair-photons. 

KTP is particularly appealing as a source for dispersion-controlled photons. As seen in Fig.~\ref{fig:gv}, it exhibits both aGVM and sGVM conditions at different frequencies. In particular, through the sGVM condition, it can be used to produce photon pairs with degenerate spectra in the highly useful telecommunications wavelength regime. In 2011, researchers at NIST presented a highly pure and spectrally degenerate telecom PDC source realised in bulk ppKTP~\cite{Gerrits2011}, demonstrating the indistinguishability of the photon pair through 95\% visibility in signal-idler Hong-Ou-Mandel interference.

To achieve the long interaction lengths necessary for narrow phasematching functions, sources in guided-wave media are essential. In addition, the tight field confinement provides significant increases in the source brightness, and the spectral and spatial degrees of freedom are largely decoupled in a waveguide. In 2011, the group of C. Silberhorn presented the first separable PDC source in a waveguide~\cite{Eckstein2011}, based on Rubidium-exchanged ppKTP. The tight field confinement contributed to a high brightness, with $\langle\hat{n}_\mathrm{PDC}\rangle\approx2.5$ photons per pulse at pump pulse energies as low as $70~pJ$, and the purity of the source was confirmed through both JSI and $g^{(2)}$ measurements. A further refinement of the source offered a signal-idler indistinguishability of around $94\%$ confirmed with Hong-Ou-Mandel interference, and a photon spectral purity of up to $86.7\%$ was obtained from interfering the photon with a classical reference field~\cite{Harder2013}. Since then, sGVM sources have been incorporated into dual-pumped Sagnac schemes to construct degenerate and highly pure photon pair sources with polarisation entanglement~\cite{jin2014pulsed,weston2016efficient}.

\subsection{The problem with side lobes}\label{sec:sinc}

\begin{figure}
\centering
\includegraphics[width=.9\linewidth]{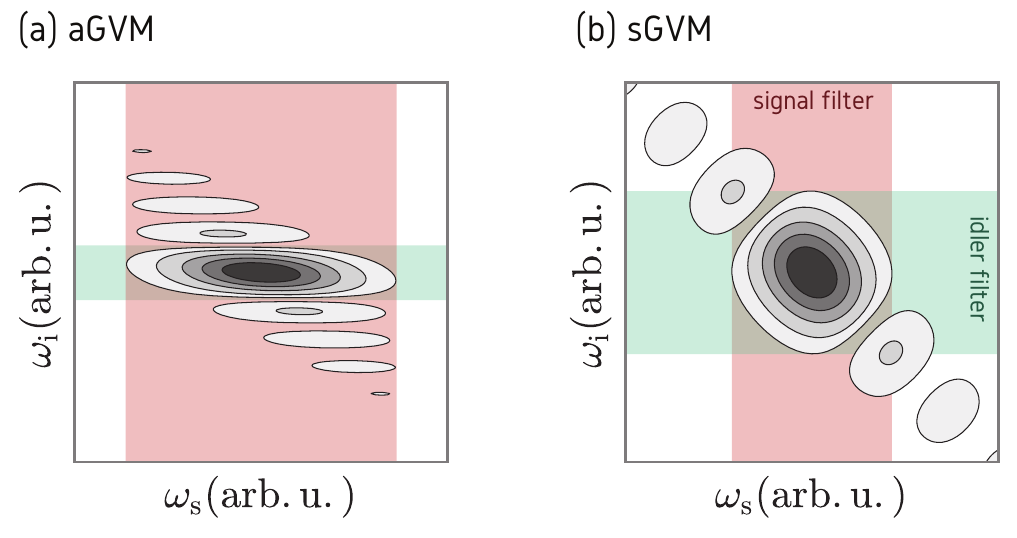}
\caption{Joint spectral amplitudes (absolute value) with standard periodic poling and filters on the individual photons. (a) In an aGVM source, the idler can be filtered to remove the side lobes and herald pure signal photons. However, filtering on the signal arm cannot be used to remove the side lobes. (b) In sGVM sources, the JSA is symmetric. Filtering either signal or idler leaves the other with a purity of about 94\%.}
\label{fig:max_purity}
\end{figure}

To put these results into context, we next consider the limitations imposed by the phasematching function in \eqref{eq:phasematching1}. In Fig.~\ref{fig:max_purity}, we plot the JSAs resulting from this phasematching function, along with possible broadband spectral filtering. It becomes immediately obvious that the side lobes of the sinc-shaped phasematching function introduce undesired frequency anti-correlations, limiting the maximum purity of heralded photons to around 86\% in the sGVM case.  With filters chosen to transmit the main peak of the JSA but block as many of the correlated side lobes as possible, it is possible to increase the source performance, but limitations are still present. In the case of aGVM depicted in Fig. \ref{fig:max_purity}(a), the idler filter can be chosen to be much narrower than the signal filter. In this example, if the idler is filtered and serves as herald, the maximum purity for the heralded signal increases to 97\%. In contrast, if the signal is filtered and serves as a herald, the heralded idler photon has a maximum purity of about 92\%. Note that this value can be increased with a larger pump bandwidth. In sGVM example shown in Fig. \ref{fig:max_purity}(b), the signal and idler photons are indistinguishable, and the filtering shown in either case leaves the other photon with a purity of about 94\% when heralded. We note that these numbers can be further increased when choosing smaller filter bandwidths at the cost of decreased heralding rates~\cite{URen2003,Meyer-Scott2017}. 

\begin{figure*}[!ht]
\centering
\includegraphics[width=1.\linewidth]{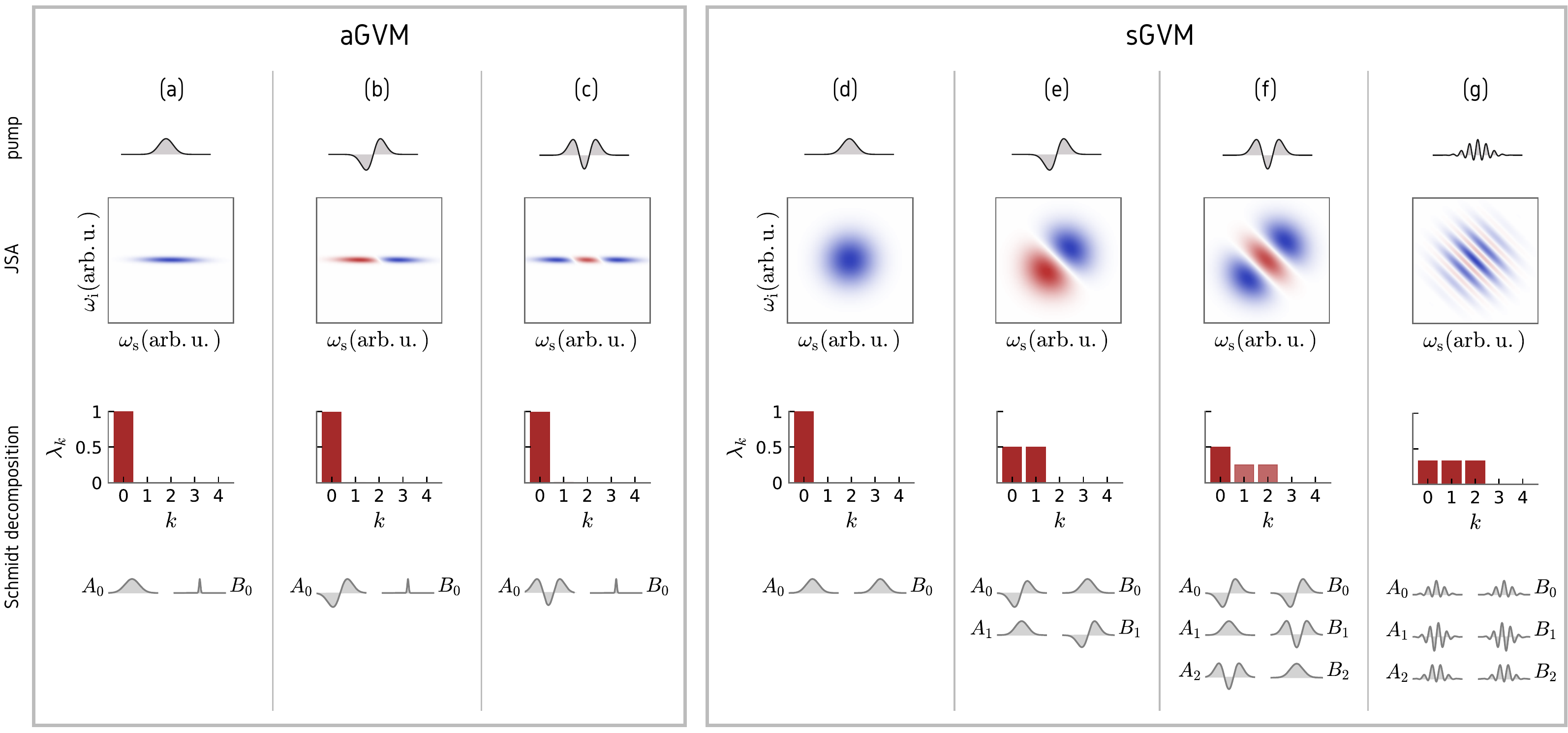}
\caption{Orchestrating Schmidt modes via group-velocity matching and pump pulse shaping. (a-c) JSAs for a PDC source with an aGVM setting. The weights of the first five Schmidt modes $\lambda_k$ are shown under each JSA. The state remains single-mode regardless of the pump shape. The only significant Schmidt modes of signal $A_0$ and idler $B_0$ photons are shown at the bottom, where we plot TM amplitudes versus frequency. The idler photon shape is invariant to the pump, while the TM of the signal photon reflects the TM of the pump field. (e-g) A sGVM PDC can be used to control the exact number of excited TMs. For example, driving the source with a first-order Hermite-Gaussian pump pulse as in (e) results in exactly two TMs. This can be extended with higher orders of Hermite-Gaussian pulses as in (f), but the different Schmidt modes are not occupied with the same probability. A balanced Schmidt-mode distribution can be achieved when the source is pumped with time-bin superpositions, as in (g).}
\label{fig:jsa_eng}
\end{figure*}

Luckily, there are elegant methods to shape the phasematching function in order to avoid the spectral filtering. These methods rely on engineering the phasematching distribution through modulation of the poling patterns and, in the case of integrated devices, tailoring the geometry of the waveguided structures. Since the phasematching function is the Fourier transform of the quasiphasematching (QPM) grating ($\chi(z)$ in \eqref{eq:phasematching0}), the nonlinearity profile along the interaction can be smoothened or \emph{apodised} to a Gaussian function by modulating the QPM grating. The first experimental demonstration of phasematching apodisation was realised by the group of M. Fejer in 2006 \cite{Huang2006}, where 13~dB suppression of the side-lobes is shown. This simple apodisation method reduces the peak efficiency and broadens the width of the phasematching function, as expected from the Fourier analysis. Apart from custom QPM gratings, the authors also investigate different waveguide geometries effective for eliminating the phasematching side-lobes. Later, many other methods were proposed and demonstrated to efficiently apodise the phasematching function, such as modulation of the poling periodicity \cite{Branzcyk2011}, modulating the poling pattern's duty-cycle \cite{Dixon2013,Chen2017}, and optimising the orientation of each domain \cite{Dosseva2016,Tambasco2016,Graffitti2017,Graffitti2017a}. These techniques grant purities in excess of 99\% without spectral filtering, opening new avenues to engineer the TM structure of PDC states by arbitrary shaping of the phasematching function.

\subsection{Controlled generation of temporal modes}
Finally, we want to highlight two possibilities to accurately control the generated PDC state beyond separability. For applications that exploit TMs as the encoding basis, the targeted generation of states with a user-defined TM structure is highly desirable. Complementary techniques arise for PDC state engineering through spectrally shaping the pump pulse in aGVM and sGVM sources, the former providing pure shaped single photons while the latter provides flexible sources for high-dimensional TM entanglement.

In the aGVM case, as seen in \eqref{eq:jsa_aGVM}, the spectrum of the idler photon is almost entirely dependent on the phasematching while the spectrum of the signal photon is dependent on the shape of the pump. By manipulating the spectral shape of the pump, the shape of the signal photon can be programmed on-the-fly, as seen in  Fig.~\ref{fig:jsa_eng}(a-c). So long as the phasematching is narrow relative to the finest features of the desired spectral shape, the JSA remains separable. This was recently demonstrated in KTP waveguides under birefringent phasematching conditions, providing high-purity shaped photons at 1411~nm~\cite{Ansari2017a}.

In contrast, PDC states that comprise a user-defined number of TMs can be generated in the sGVM configuration. Again, this is achieved by spectral shaping of the pump pulses. One example of this is a PDC driven by a pump pulse with a first-order Hermite-Gaussian spectrum \cite{Brecht2015}, as depicted in Fig. \ref{fig:jsa_eng}(e). In this case, the generated state is a TM Bell-state of the form
\begin{equation}
|\psi\rangle_\mathrm{Bell} = \frac{1}{\sqrt{2}}\left(|0\rangle_\mathrm{s}|1\rangle_\mathrm{i}+e^{\imath\varphi}|1\rangle_\mathrm{s}|0\rangle_\mathrm{i}\right),
\end{equation}
where $|0\rangle_j$ ($|1\rangle_j$) labels the $j$ photon occupying a Gaussian (first-order Hermite-Gaussian) spectrum and $j$=(s,i). To add additional TMs to this state, it is sufficient to increase the order of the Hermite-Gaussian spectrum of the pump pulse, which is easily achieved with conventional pulse shaping~\cite{weiner2000femtosecond}. Although this provides a state with finite number of Schmidt modes, the generated TMs are generally not equally occupied (i.e. they can have different $\sqrt{\lambda_k}$)~\cite{Brecht2015}, and thus the generated TMs are not maximally entangled. Another alternative pump shape to control the Schmidt modes is a superposition of time bins or, equivalently, cosine functions in the frequency domain, as shown in Fig. \ref{fig:jsa_eng}(g) \cite{Patera2012}. This provides a flexible and versatile source that generates maximally entangled states with an arbitrary dimension without the need for changing any hardware.

As the last remark in this section, we want to point out that the theoretical description of the PDC process presented here, using the first-order perturbation theory, is only valid when the process is weakly pumped (also referred to as the low-gain regime)~\cite{Christ2013, Quesada2015}. A full description of such nonlinear optical processes requires the time-ordered treatment of the involved Hamiltonians and consideration of the presence of multi-photon components. In the high-gain regime (with intense pump powers and PDC mean photon numbers $\gg~1$), the time-ordering leads to significant changes of the Schmidt modes and the respective squeezing in each mode. Despite this, in the high-gain regime it is possible to generate bright squeezed states which are interesting to study a range of quantum phenomena at mesoscopic scales~\cite{lemieux2016engineering,Harder2016,Harder2017}.

To conclude, PDC state engineering is now at a point where we can exert close-to arbitrary control over the TM structure of the generated state. This brings into reach the realisation of TM based QIP applications and provides us with a very clean laboratory system for the generation of Hilbert spaces with well-defined dimensions.

\section{Manipulation and measurement of temporal modes}\label{sec:three}

With a variety of sources available for both pure and entangled TM-encoded photons, the next piece of the complete TM-based QIP toolbox is a quantum device capable of accessing a TM out of a multimode input. In other words, we require a special quantum-mechanical beam splitter that operates on a customisable basis of TMs. A promising tool to build such a device is engineered frequency conversion.

Frequency conversion (FC) has been recognised as means to translate the central frequency of a photonic quantum state while preserving its non-classical signatures. The first proposal in 1990 considered the frequency-translation of squeezed states of light~\cite{Kumar1990}. Different experiments have since confirmed that FC retains quadrature squeezing~\cite{Huang1992,Vollmer2014,Baune2015,Liu2015}, quantum coherence and entanglement~\cite{Tanzilli2005,Honjo2007,DeGreve2012,ramelow2012polarization}, anti-bunching of single photons~\cite{Zaske2012, Ates2012}, and non-classical photon correlations~\cite{Rakher2010,McGuinness2010}. Since FC can be highly efficient~\cite{Vandevender2004,VanDevender2007,Pelc2011}, it provides a useful tool for improved detection schemes~\cite{Albota2004,Roussev2004,Langrock2005,Thew2006,Legre2007} and an interface for dissimilar nodes in future quantum networks~\cite{Giorgi2003, Ding2010, Takesue2010, Ikuta2011, Curtz2010, Ikuta2014, Kuo2013, Cheng2015a,Steinlechner2016,Ruetz2017,maring2017photonic}. 

However, there is more to frequency conversion. In 2010, Raymer and co-workers proposed an interpretation of FC as a two-colour beam splitter~\cite{Raymer2010}, enabling for example Hong-Ou-Mandel interference~\cite{Hong1987} of photons of different colour. If the FC is set to 50\% efficiency, and if two monochromatic photons which are centred at the two linked frequencies (red and blue) are sent into the process, simultaneous SFG/DFG occurs and both photons will exit the FC either at the blue frequency or the red frequency. The conversion process links the two frequency bands in a beam splitter fashion, as has been demonstrated with single-photon signals exhibiting Ramsey interference~\cite{Clemmen2016} and two-colour Hong-Ou-Mandel interference~\cite{kobayashi2016frequency}.

The proposal of Ref.~\cite{Raymer2010} also considers the case of spectrally broadband FC, where a specific input frequency $\oin$ is mapped to a plethora of output frequencies $\oout$ and vice versa, as determined by the Heisenberg-picture Bogoliubov transformations
\begin{equation}
\begin{split}
\hat{a}^\dagger(\oin)\longmapsto&\int d\oin'G_\mathrm{aa}(\oin,\oin')\hat{a}^\dagger(\oin')\\
&+\int d\oout'G_\mathrm{ac}(\oin,\oout')\hat{c}^\dagger(\oout'),
\end{split}\end{equation}
\begin{equation}\begin{split}
\hat{c}^\dagger(\oout)\longmapsto&\int d\oin'G_\mathrm{ca}(\oout,\oin')\hat{a}^\dagger(\oin')\\
&+\int d\oout'G_\mathrm{cc}(\oin,\oout')\hat{c}^\dagger(\oout').
\end{split}\end{equation}
Here, $\hat{a}^\dagger$ and $\hat{c}^\dagger$ are creation operators in the input and frequency-converted output modes, respectively, and the $G_{ij}$ are Green's functions that describe the mapping between the two. By applying a Schmidt decomposition to the Green's functions, an interpretation of broadband FC as a beam splitter that links sets of input TMs to output TMs becomes apparent~\cite{Raymer2010}. Similar to PDC, this process generally will be multimode. 

Inspired by the previously outlined work in PDC engineering, the mode structure of FC can be tailored through dispersion engineering. It turns out that a configuration that is similar to asymmetric group-velocity matching facilitates single-mode operation: when the input signal propagates through the nonlinear medium at the same velocity as the bright pump but the output is group-velocity mismatched, one specific TM is selected and converted to the output frequency, while all other TMs are simply transmitted~\cite{Eckstein2011a}. The single-mode FC has been dubbed the \textit{quantum pulse gate} (QPG) to reflect that it selects, or gates, one broadband TM. The reversal of this process, when the output light shares the group velocity of the pump, has been proposed as a TM shaper~\cite{Brecht2011a}.

\begin{figure}[t]
\centering
\includegraphics[width=.9\linewidth]{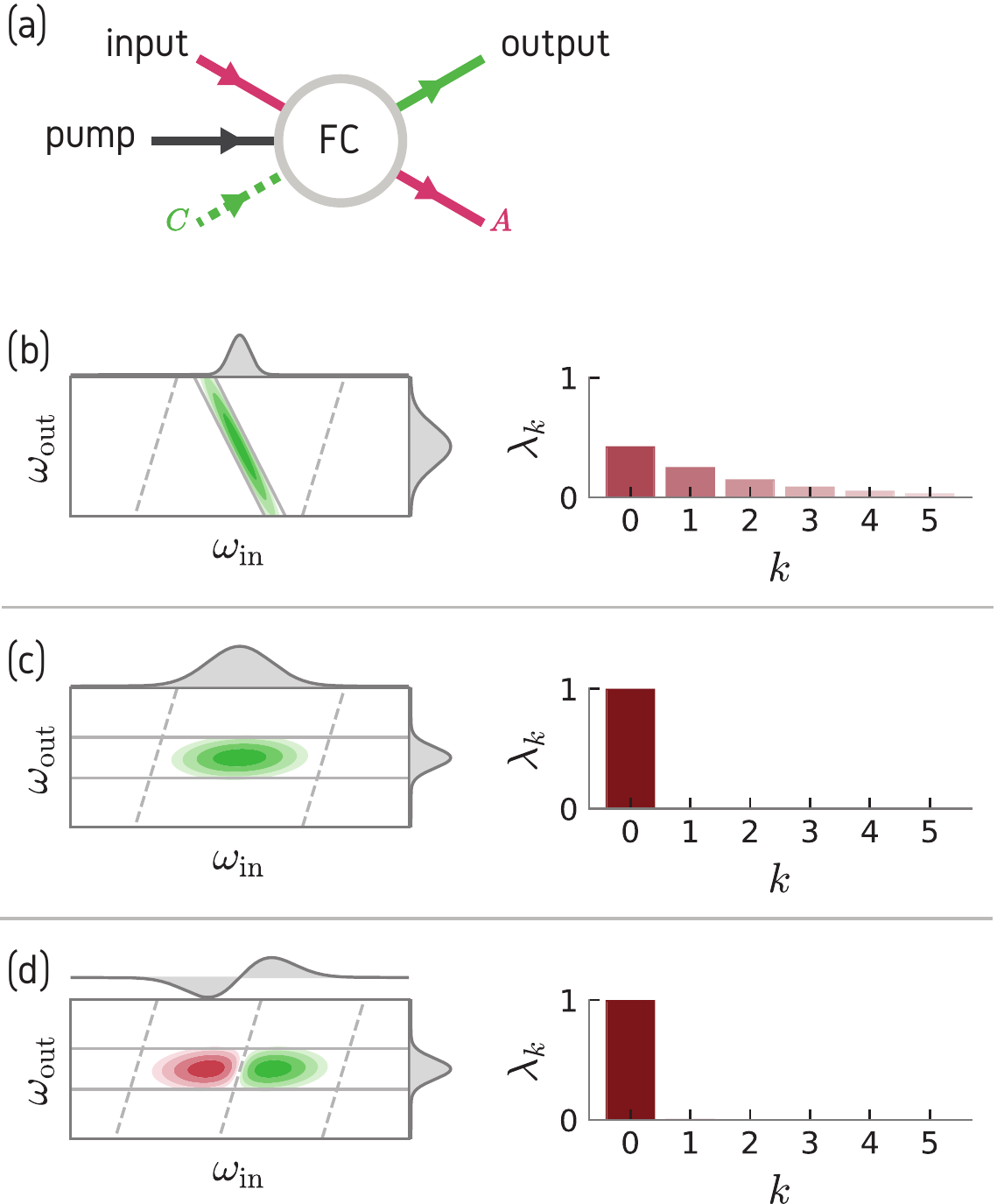}
\caption{Frequency conversion process and its transfer function. (a) Outline of a general frequency conversion process with pump, input and output fields. (b to d) Sum-frequency conversion transfer functions $F(\oin, \oout)$ with its marginal distributions (left) and its first few Schmidt coefficients $\sqrt{\lambda_k}$. (b) A non-engineered SFG with significant frequency correlations and a $K\approx3.7$. (c) and (d) present a tailored SFG process with aGVM condition with pump functions $\alpha(\oout-\oin)$ of Gaussian and first-order Hermite-Gauss, respectively, and a $K\approx1.01$.}
\label{fig:qpg_tf}
\end{figure}

In the following we briefly outline the QPG formalism. The interaction Hamiltonian that describes a general FC process is given by
\begin{equation}
\hat{H}_\mathrm{int}=\theta\int d\oin\,d\oout F(\oin, \oout)\hat{a}(\oin)\hat{c}^\dagger(\oout) + \mathrm{h.c.},
\end{equation}
where $\hat{a}$ and $\hat{c}$ are annihilation operators in the input and upconverted modes, respectively, and $\theta$ is a coupling of the process incorporating the power of the QPG pump and the strength of the material nonlinearity. The \textit{transfer function} $F(\oin, \oout)$ describes the mapping from input to output frequencies, equivalent in the low-efficiency regime to the Green's function $G_{\mathrm{RB}}(\oin, \oout)$ and analogous to the JSA in PDC processes. The transfer function, as in the case of PDC, is a product of pump amplitude and phasematching \begin{equation}F(\oin, \oout)=\alpha(\oout - \oin)\phi(\oin, \oout).\label{eq:transfunc}\end{equation} Similar to PDC, we can apply a Schmidt decomposition to the mapping function and define our operators in the TM basis (compare \eqref{eq:jsa} - \eqref{eq:pdc_schmidt}), obtaining
\begin{equation}
\hat{H}_\mathrm{int} = \theta\sum_{k=0}^\infty\sqrt{\lambda_k}\hat{A}_k\hat{C}_k^\dagger + \mathrm{h.c.},
\label{eq:fc_schmidt}
\end{equation}
with $\sum_k\lambda_k=1$. Despite the similarity to the Schmidt decomposition of the PDC state as formulated in Eq.~(\ref{eq:pdc_schmidt}), there is a fundamental difference in the meaning of the decomposition and the Schmidt modes in each case. While the PDC decomposition expresses the modes of a \textit{state}, in the case of the FC we have a SFG \textit{operation}. The Hamiltonian in Eq.~(\ref{eq:fc_schmidt}) generates operator transformations
\begin{align}
\hat{A}_k &\rightarrow \cos(\sqrt{\lambda_k}\theta)\hat{A}_k + \sin(\sqrt{\lambda_k}\theta)\hat{C}_k,\\
\hat{C}_k &\rightarrow \cos(\sqrt{\lambda_k}\theta)\hat{C}_k - \sin(\sqrt{\lambda_k}\theta)\hat{A}_k.
\end{align}
These can be interpreted as $k$ independent beam splitters with reflectivities $\sin^2(\sqrt{\lambda_k}\theta)$, which connect the input $\hat{A}_k$ to an output $\hat{C}_k$.

As previously derived for PDC, the phasematching function can be written in terms of the group-velocity mismatch, $\Delta{k}(\oin,\oout)$. Assuming that the nonlinear medium is periodically poled to ensure phasematching at the centre frequencies, this phase mismatch can be written to first order in analogy to ~\eqref{eq:delKpdc1stord} as \begin{equation}\Delta k(\oin,\oout) \approx (u^{-1}_{\mathrm{in}}-u^{-1}_{\mathrm{p}})\oin - (u^{-1}_{\mathrm{out}}-u^{-1}_{\mathrm{p}}) \oout.\label{eq:delKsfg1stord} \end{equation} For the case of aGVM where the input signal propagates at the same velocity as the pump ($u_{\mathrm{in}}=u_{\mathrm{p}}$), the first-order phasematching function is only dependent on the upconverted frequency ${\phi(\oin,\oout)\approx\tilde\phi(\oout)}$. If the phasematching is spectrally narrow enough that the output frequency spread is negligible compared to the input, the contribution of the pump field is approximately dependent on only the frequency of the input field, $\alpha(\oout-\oin)\approx\tilde{\alpha}(\oin)$. If these approximations hold, the transfer function can be rewritten simply as
\begin{equation}F(\oin,\oout) \approx \tilde{\alpha}(\oin) \tilde\phi(\oout).\end{equation} As the phasematching function tightens, the transfer function becomes more and more separable, as illustrated in Fig.~\ref{fig:qpg_tf}(c,d).

For a separable transfer function, the Schmidt decomposition yields only one single non-zero Schmidt coefficient and the interaction Hamiltonian reduces to the desired QPG Hamiltonian,
\begin{equation}
\hat{H}_\mathrm{QPG} = \theta\hat{A}_0\hat{C}_0^\dagger+\mathrm{h.c.}
\label{eq:qpg_schmidt}
\end{equation}
and we obtain the following operator transformations
\begin{align}
\hat{A}_0 &\rightarrow \cos(\theta)\hat{A}_0 + \sin(\theta)\hat{C}_0,\\
\hat{C}_0 &\rightarrow \cos(\theta)\hat{C}_0 - \sin(\theta)\hat{A}_0,\\
\hat{A}_{k} &\rightarrow \hat{A}_k\text{ for }k\neq0,\\
\hat{C}_{k} &\rightarrow \hat{C}_k\text{ for }k\neq0.
\end{align}
Hence, the ideal QPG selects one single input TM and converts it to an output TM with an efficiency of $\sin^2(\theta)$, while all orthogonal TMs pass through the QPG unconverted and undisturbed. The selected input TM $\hat{A}_0$ is defined by the shape of the bright pump pulse that drives the conversion ($\tilde{\alpha}(\oin)$), whereas the shape of the output TM $\hat{C}_0$ is given by the envelope of the phasematching function ($\tilde\phi(\oout)$)~\cite{Eckstein2011a,Brecht2011a}. By shaping the spectral amplitude and phase of the QPG pump pulse, the mode selected by the QPG can be adapted on-the-fly. While most works have motivated the QPG towards Hermite-Gauss TMs, it can also be set to select arbitrary superpositions as well as entirely different mode bases (e.g. time or frequency bins) by reshaping the pump pulse. While other group-velocity conditions exist which enable nearly single-mode sum-frequency generation, the aGVM case outlined here has been shown to be optimal~\cite{Reddy2013}.

\begin{figure}
\centering
\includegraphics[width=.91\linewidth]{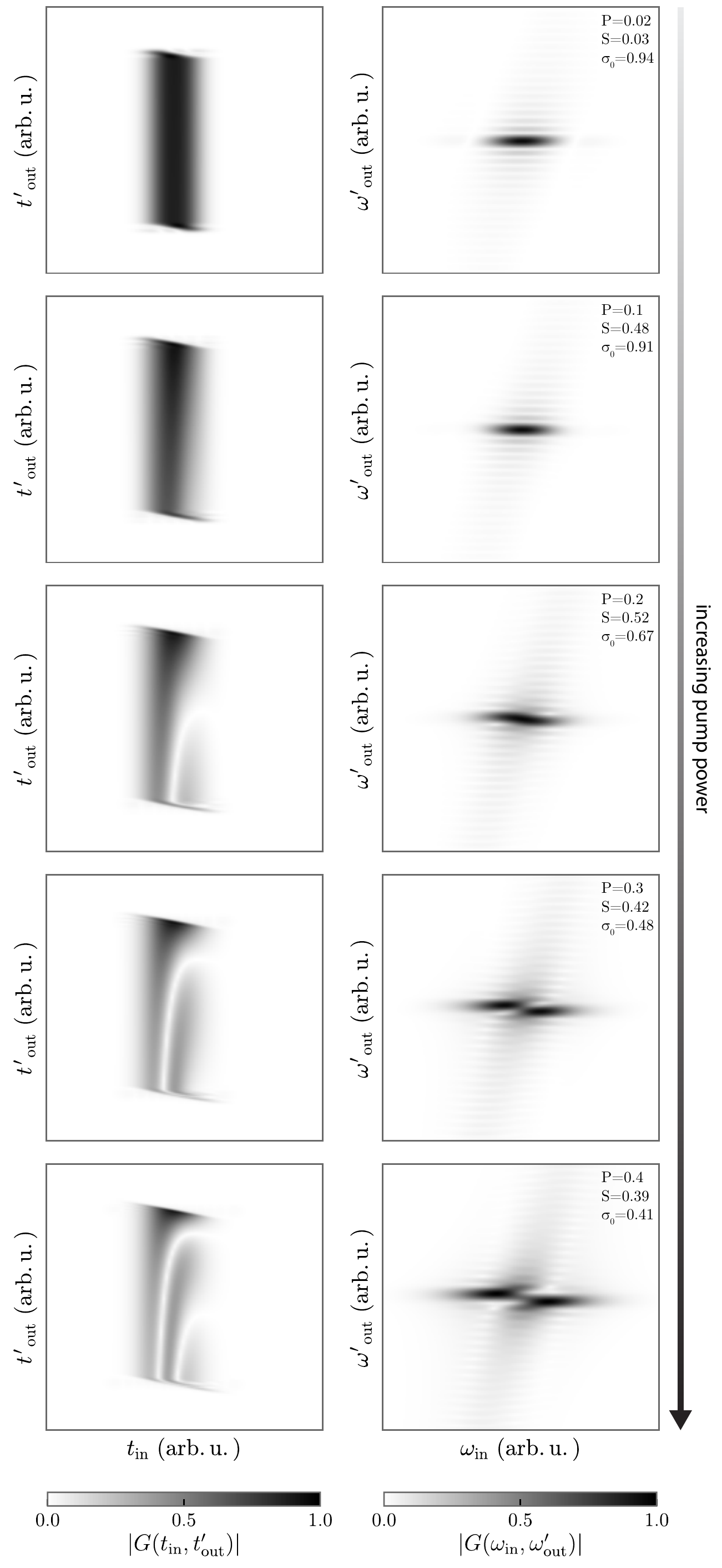}
\caption{Absolute value of the temporal (left) and spectral (right) transfer functions for broadband frequency conversion. The left column shows the mapping from input times $t_\mathrm{in}$ to output times $t_\mathrm{out}$ for increasing pump powers (top to bottom), corresponding to increasing conversion efficiencies. The relative pump energy $P$, selectivity $S$, and separability $\sigma_0$ are printed on top-right corner of each row. This leads to simultaneous forward and backward conversion, which is reflected by the oscillations in the mapping function. The functions were calculated by numerically solving the Heisenberg equations for the input and output field operators. The right column shows the respective spectral mapping functions. It can be seen that the general shape of the function broadens and that additional correlations are introduced for stronger pump powers. These correlations do not show up in a perturbative approach.}
\label{fig:green}
\end{figure}

Although ideal QPG operation as described in Eq. (\ref{eq:qpg_schmidt}) requires perfect GVM between the pump and input, one can still realise a nearly single-mode QPG if the group-velocity mismatch is small, with respect to the temporal width of each field. To compare different scenarios, we redefine the group-velocity mismatch contrast, was introduced in Eq. (\ref{eq:gvmc}), as
\begin{equation}
\xi = \frac{ u^{-1}_{\mathrm{in}} - u^{-1}_{\mathrm{p}} }{ u^{-1}_{\mathrm{out}} - u^{-1}_{\mathrm{p}} }.\label{eq:SFGaGVM}
\end{equation}
An aGVM condition between the pump and input fields means $\xi \to 0$. This definition can help us to study the feasibility of building a QPG in different non-linear materials with different dispersion properties, which will be discussed in the next section.

More detailed studies followed this first proposal for a QPG, which focused in particular on the behaviour of a QPG as a function of conversion efficiency. In this context, implementations based on both four-wave mixing and SFG were investigated \cite{Reddy2013}. The figure of merit that was defined is the so-called \textit{selectivity} $S$ of the QPG, which is defined as
\begin{equation}
S = \eta_0\cdot\frac{\eta_0}{\sum_{k=0}^\infty\eta_k}\leq1,
\end{equation}
where $\eta_k=\sin^2(\sqrt{\lambda_k}\theta)$ is the conversion efficiency for the $k$-th TM. The selectivity measures both the single-modedness of the QPG and the conversion efficiency for this mode. 

An ideal QPG operates on only one TM and converts this mode with unit efficiency ($S=1$). In a more realistic scenario, the QPG becomes multimode when approaching high conversion efficiencies, owing to non-perturbative interaction dynamics often referred to as \textit{time-ordering} effects in the quantum context~\cite{Christ2013,Reddy2013,Quesada2014}. For a single QPG, a maximum selectivity of $S\approx83\%$ has been determined~\cite{Reddy2013}. Fig.~\ref{fig:green} shows the change in the transfer functions for increasing pump powers~\cite{Reddy2013}.

In Ref.~\cite{Reddy2014}, Reddy et al. proposed a scheme to overcome this limitation, dubbed \textit{temporal-mode interferometry}. Using two QPGs in a Mach-Zehnder-like configuration, they show it is possible to achieve selectivities approaching unity. In this scheme, two QPGs are operated at 50\% conversion efficiency---similar to two balanced beam splitters---and the phases between the two QPGs are adjusted such that interference leads to complete conversion of the targeted input TM. Since each QPG operates at a moderate conversion efficiency, the individual processes are still close-to single-mode and an overall selectivity of more than 98\% can be achieved. 

Despite this advance, simultaneously achieving high efficiency and isolating orthogonal modes is a significant experimental challenge. In scenarios where the QPG is used for temporal-mode reconstruction and measurement, efficiency may not be the dominant concern. Instead, one might simply need to know how well the upconverted signal identifies the presence of the target TM. To isolate this criterion, often the \textit{separability} $\sigma_j$ for a given mode $j$ among a $d$-dimensional basis is quantified, defined as~\cite{manurkar2016multidimensional} \begin{equation}\sigma_j=\frac{\eta_j}{\sum_{k=0}^d\eta_k} \leq 1 \label{eq:separability}\end{equation} This quantifies how well the QPG isolates a single mode from a mixture irrespective of incomplete conversion. Additionally, oftentimes the suppression or \textit{extinction ratio} for mode $j$ is reported~\cite{Brecht2014,Kowligy2014}, \begin{equation}\mathrm{E.R._j\,(dB)} =10\log_{10}\frac{\eta_j}{\max_{k\neq j}\eta_k}, \end{equation} which defines to which extent the QPG suppresses signals from modes orthogonal to the target mode.

\section{Experimental progress on TM selection}\label{sec:four}
\begin{figure}
\centering
\includegraphics[width=.8\linewidth]{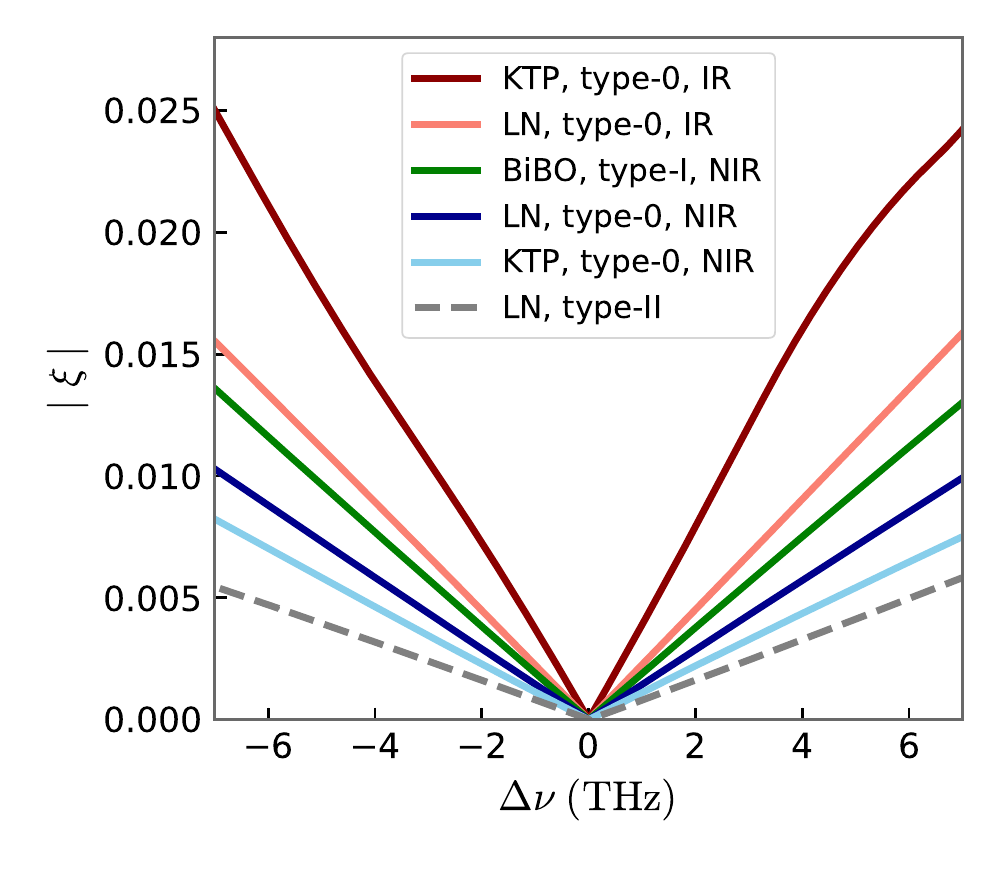}
\caption{The group-velocity mismatch contrast $\xi$ (such that 0 is perfectly matched) for processes in lithium niobate (LN) waveguides, potassium titanyl phosphate (KTP) waveguides, and bulk bismuth borate (BiBO), as the input signal is detuned from the optimal group-velocity matching. The grey dashed line corresponds to the type-II process in LN, where GVM is found for a 1550~nm signal, 875~nm pump, and 560~nm upconverted~\cite{Brecht2014}. All other processes have degenerate signal and QPG pump for group-velocity matching, and IR (NIR) corresponds to 1550~nm (800~nm) signal and QPG pump. Signal detuning or noncollinear geometry is necessary in all cases except for type-II LN to overcome the second harmonic of the QPG pump.}
\label{fig:gvm}
\end{figure}

In this section, we provide an overview of experimental work on temporal-mode-selective devices built with pulse shaping and dispersion engineering. To start, it is imperative to find nonlinear materials and interactions that satisfy the aGVM conditions, i.e. minimise $|\xi|$ in~\eqref{eq:SFGaGVM}. This condition can be met for SFG processes in multiple materials, as mapped out in Fig.~\ref{fig:gvm}. In particular, it naturally occurs near degeneracy in materials with type-0 or type-I phasematching conditions (i.e. where the QPG pump and input have the same polarisation and approximately the same frequency). However, in these near-degenerate configurations, the second harmonic of the QPG pump adds a strong source of phasematched background noise for single-photon operation, and suppressing it by detuning the signal from degeneracy quickly degrades the mode selectivity of the device, as seen in the rising $\xi$ values in Fig.~\ref{fig:gvm}. To operate with ``perfect'' group-velocity matching, specific conditions can be found in type-II or frequency-nondegenerate configurations. For example, in z-cut lithium niobate, a 1550~nm ordinarily polarised input signal may interact with a 875~nm extraordinarily polarised QPG pump to produce an ordinarily polarised upconverted signal in the green range of the visible spectrum~\cite{Eckstein2011a,Brecht2014}. Since the SHG process for the QPG pump is both phase mismatched and in the blue range, the upconverted signal can be effectively isolated at the optimal GVM wavelength. However, the type-II nonlinear strength is considerably weaker than the type-0, necessitating stronger pump fields.

While broadband temporal modes find a natural use in quantum applications, similar concepts have been proposed and explored for classical communications. By taking a broad flat-top optical pulse and manipulating its spectral phase with a pulse shaper, one can generate sets of orthogonal pulses based on, for example, Hadamard codes. If a decoder applies the correct decoding phase sequence, the ultrashort pulse becomes Fourier limited once more, with a commensurate increase in peak power~\cite{salehi1990coherent}. This concept can be merged with dispersion-engineered sum-frequency generation to enable ultrashort-pulse code-division multiple access. If a broadband pulse is sent through a long nonlinear crystal for second-harmonic generation (SHG), and the crystal is group-velocity mismatched such that the SHG light walks off from the input light, the second harmonic will be temporally lengthened and spectrally narrowed. If a frequency-dependent phase is applied to the pulse, it will only be efficiently frequency doubled if the phase is symmetric. If two users each have access to half of the spectral bandwidth of an ultrashort pulse, the pulse will cease to upconvert in this medium if they apply orthogonal phase codes~\cite{zheng2000spectral,zheng2001low}. This effect is due to interference within the broadband pulse structure and enabled by the group-velocity walkoff in the nonlinear medium. This scheme was demonstrated by the group of A.~M.~Weiner using a 20-mm long bulk PPLN sample with a broad input pulse at telecommunications wavelength split into 16 channels. The SHG from mismatched codes exhibited an extinction ratio of over 27~dB when filtering the central frequency component~\cite{zheng2000spectral}. Using entangled photon pairs to supply the same effective spectral narrowing as the group-velocity mismatched SHG, analogous encoding schemes have been demonstrated with biphoton upconversion~\cite{Lukens2014}.

\begin{figure}
\centering
\includegraphics[width=1.\linewidth]{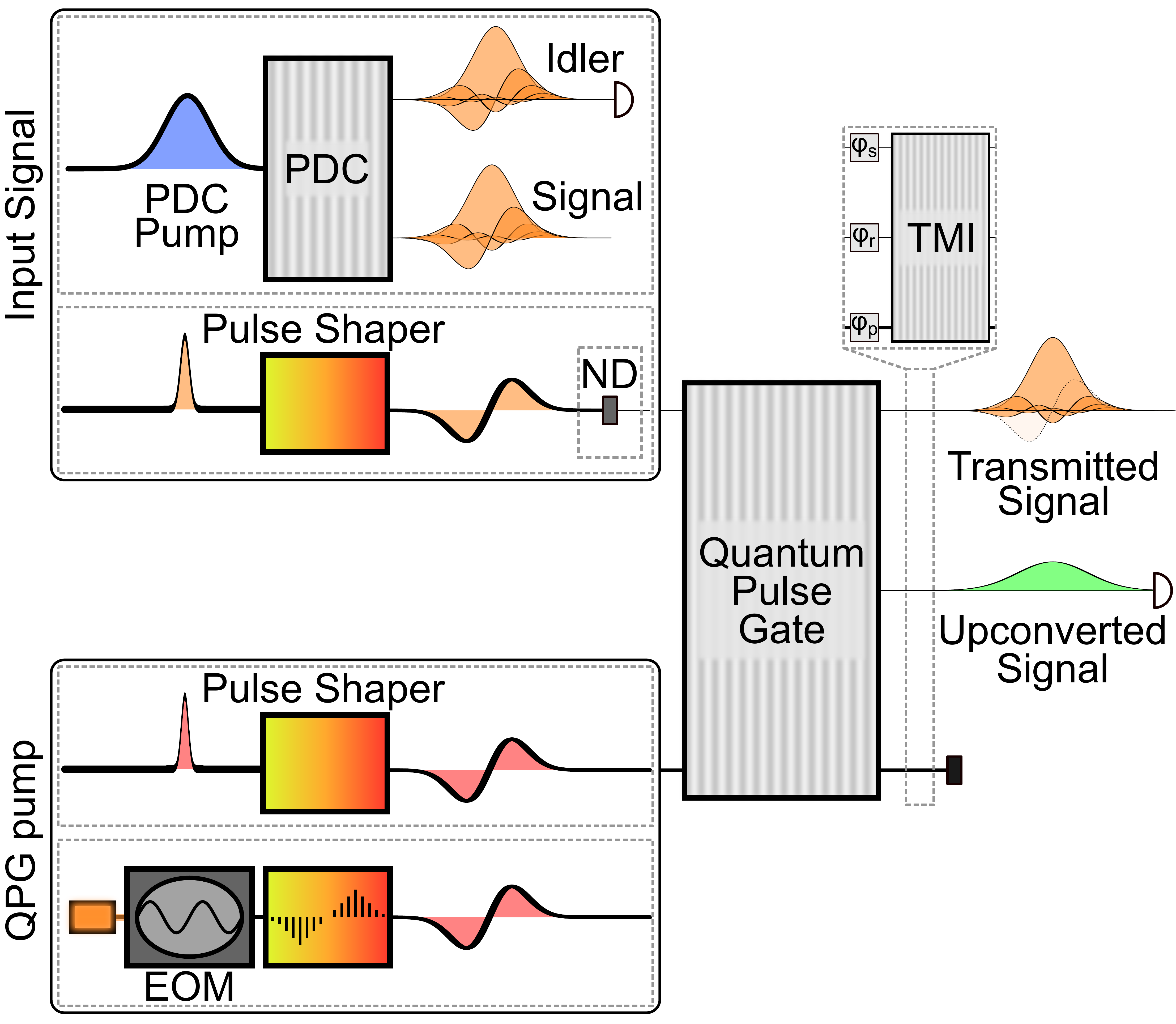}
\caption{Generic experimental situation for a quantum pulse gate. A TM-encoded single photon or weak coherent state is prepared through PDC or through shaping a spectrally broad input pulse and attenuating with a neutral density (ND) filter. A strong QPG pump is prepared using similar pulse shaping methods, or through electro-optic modulation (EOM) of a strong cw laser to produce a frequency comb which is modulated in a tooth-by-tooth fashion by a pulse shaper~\cite{Kowligy2014}. The two are mixed in a group-velocity-matched $\chi^{(2)}$ waveguide, and the upconverted signal in the register mode is measured. For temporal-mode interferometry (TMI)~\cite{Reddy2014}, the QPG is split into two 50\% efficient steps with phase shifts in between.}
\label{fig:expsetup}
\end{figure}

\begin{figure}
\centering
\includegraphics[width=1.\linewidth]{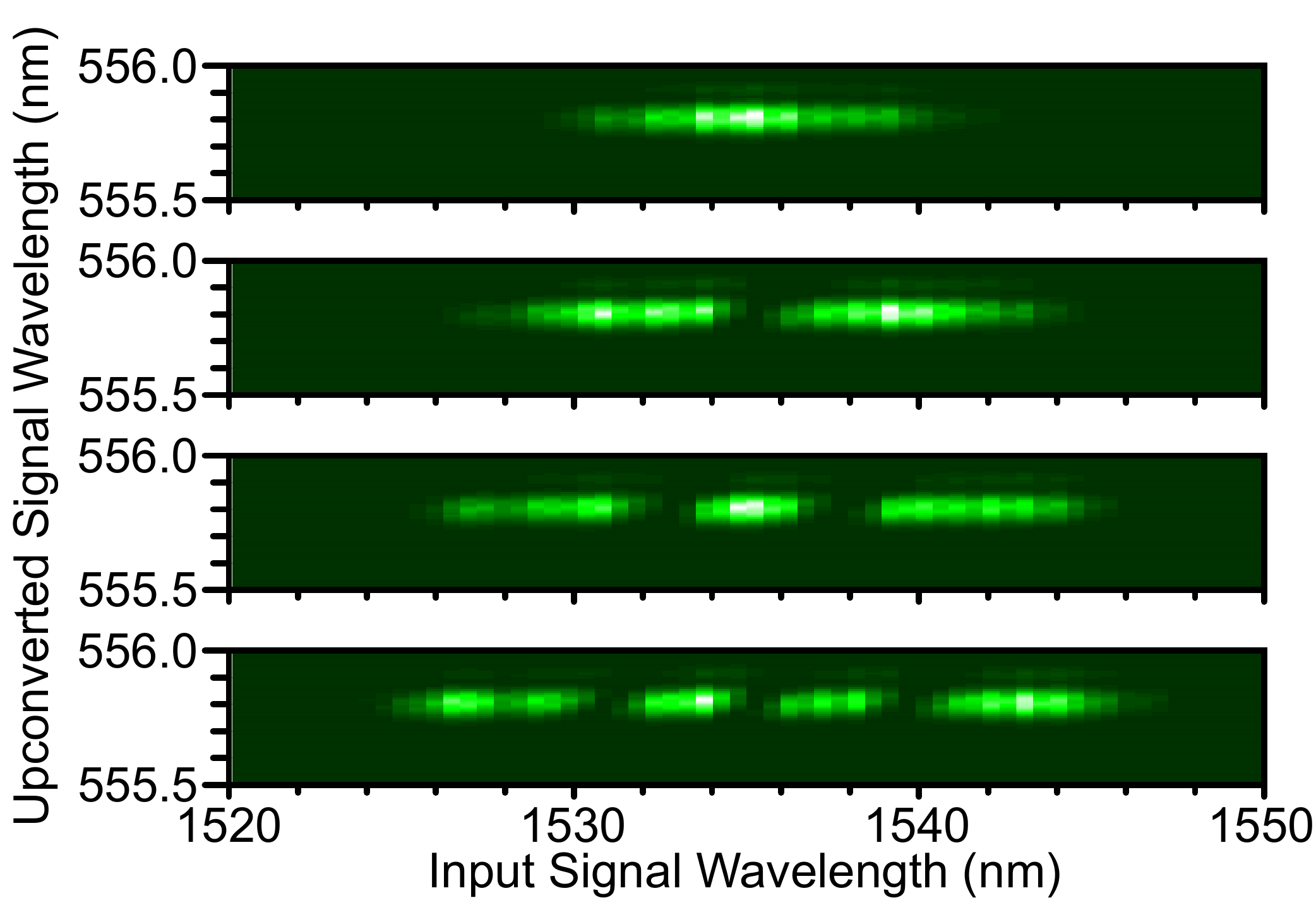}
\caption{Experimental spectral-intensity transfer functions for the first four Hermite-Gaussian temporal modes (top to bottom), as measured in the experimental apparatus of Ref.~\cite{Ansari2017}. The QPG in question was built from a 17-mm long PPLN waveguide phasematched for a type-II interaction (875~nm + 1540~nm to 555.7~nm), with the group-velocity matching necessary to produce highly separable SFG transfer functions.}
\label{fig:exptransfuncs}
\end{figure}

Recent realisations of the QPG allow for the analysis and reconstruction of the temporal modes of distant single-photon level pulses. These experiments can generally be described by the apparatus of Fig.~\ref{fig:expsetup}. In the group of C.~Silberhorn, a quantum pulse gate was constructed using a type-II interaction in titanium-indiffused PPLN waveguides with short poling periods (4.4~$\mu$m)~\cite{Brecht2014}, where an orthogonally polarised and group-velocity matched telecom (1535~nm) input signal and a Ti:Sapphire (875~nm) QPG pump mix to produce a signal in a green (550~nm) upconverted beam. The broad GVM of this process allows it to be used for sub-picosecond pulses (approximately 300~fs FWHM), with the selected mode exactly matching the spectral profile of the QPG pump in the low-efficiency regime, as seen in Fig.~\ref{fig:exptransfuncs}. In Ref.~\cite{Brecht2014}, an efficiency of nearly 88\% was observed for the primary Gaussian mode with a single-photon-level coherent state input, with a demonstrated extinction ratio of approximately 7 dB, limited by the resolution of the pulse shaper. With improved QPG pump pulse shaping, this experiment was extended to measure PDC photons from a spectrally pure source with an extinction ratio of 12.8~dB and shaped coherent laser light with an extinction ratio of over 20~dB, although with a greatly reduced conversion efficiency (approximately 20\%)~\cite{Ansari2016}. Experimental SFG transfer functions using this system can be seen in Fig.~\ref{fig:exptransfuncs}.

An approximate approach to mode-selective measurement without strict group-velocity matching was later put forth by Y.-P.~Huang and P.~Kumar~\cite{Huang2013}. Although the optimal mode-selective frequency conversion configuration has been shown to be group-velocity matched~\cite{Eckstein2011a,Reddy2013}, they found that reasonably single-mode frequency conversion could be realised through numerically optimised pump shaping so long as the bandwidth of the phasematching function is significantly narrower than the bandwidth of the pump. By generating a 20~GHz pulse train through electro-optically modulating a strong CW laser, Kowligy et al. produced a 17-element frequency comb for both the input signal and QPG pump, with each tooth individually addressable in phase and amplitude. With this scheme, they were able to experimentally demonstrate efficiencies near 80\% and 8~dB extinction ratios using a 6~cm type-II PPKTP waveguide~\cite{Kowligy2014}. In follow-up work, they reverted to a nearly group-velocity matched configuration using near-degenerate type-0 SFG in a 52~mm PPLN waveguide with input signals around 1550~nm. Applying their waveform generation and numerical optimisation to this situation, they were able to demonstrate efficiencies above 75\% for a four-dimensional Hermite-Gaussian alphabet with separabilities above 65\% and as high as 87\% for picosecond-scale Gaussian pulses~\cite{manurkar2016multidimensional}. These results have been extended to novel mode-selective pulse-shaping schemes based on over-conversion in SFG~\cite{manurkar2017programmable} and demonstrations of mode-selective upconversion with efficiencies and selectivities high enough to outperform time-frequency filtering for signal isolation~\cite{Shahverdi2017}.

In the low-efficiency regime, the spectral shape prepared for the pump pulse corresponds exactly to the temporal mode selected by the QPG. In the high-efficiency regime, this first-order treatment breaks down due to the time-ordering effects outlined in the previous section and Fig.~\ref{fig:green}~\cite{Christ2013,Quesada2014,Reddy2014}. D.~V.~Reddy and M.~G.~Raymer have investigated this regime with a QPG based on a 5-mm PPLN waveguide phasematched for a type-0 interaction between an 812~nm input signal, an 821~nm QPG pump, and a 408~nm register (output) mode~\cite{Reddy2017a,reddy2017ramseyexp}. By operating with nearly degenerate pump and signal, the group-velocity mismatch between the two red modes is negligible compared to the violet upconverted mode, and the type-0 PPLN interaction provides an extremely high nonlinearity. This allowed them to saturate the QPG efficiency at reasonable QPG pump powers (85\% with 3.5~mW at 76~MHz with 500-fs pulses)~\cite{Reddy2017a}. They also confirmed numeric predictions that, in the high-efficiency regime, greater conversion efficiencies and mode selectivities can be reached with QPG pump shapes that differ from their analytically calculated low-efficiency regime counterparts.

With 50\% conversion efficiency, enhanced mode selectivity is possible through temporal mode interferometry (TMI), where phase reshaping between two 50\% efficient QPGs suppresses higher-order corrective terms~\cite{Reddy2014,Reddy2015,Quesada2016}. By passing through the same waveguide twice (necessary to ensure identical phasematching conditions), Reddy and Raymer were able to show mode-selective Ramsey interference with enhanced efficiency and mode selectivity relative to numerically calculated single-stage expectations~\cite{reddy2017ramseyexp}. This enhancement was present using the analytic low-efficiency-regime QPG pump mode shapes, removing the need for efficiency-dependent numerical optimisation.

\subsection{Mode selection in quantum memories}

A further possibility to manipulate TMs is by tailored light-matter interactions in single-mode quantum memories, in particular Raman ensemble memories. Here, the optical light field interacts with an ensemble of atoms with a $\Lambda$ energy level configuration. A strong control pulse drives a two-photon Raman transition, which maps the addressed input TM onto a so-called spinwave, which can be transferred back into an optical field by applying another strong control pulse. Similar to a QPG, the underlying equations describing this interaction can be cast into the form of a broadband beam splitter, where the shape of the strong control pulse determines the TMs that are stored and retrieved \cite{Nunn2007}. In contrast to QPGs, quantum memories give access to a wide range of accessible spectral bandwidths ranging from few MHz up to THz, depending on the physical system used to realise the memory. Recent results have shown the potential usefulness of these types of memories for the storage and manipulation of multimode quantum frequency combs \cite{Zheng2015}, and the frequency and bandwidth conversion of photons \cite{Fisher2016,Bustard2017}. By performing a process tomography, the group of I.~Walmsley has demonstrated the single-TM operation of a Raman memory \cite{Munns2017}. Similar to the single-stage QPG, the Raman memory shows a degrading single-modedness with increasing efficiency. One way around this problem is to place the memory inside a cavity, which enables both high efficiency and mode-selectivity simultaneously~\cite{Nunn2017}.

\subsection{Multimode manipulations with sum-frequency generation}

While group-velocity engineered waveguides and mode-selective interfaces are powerful tools, by definition they are unable to reshape the structure of multimode fields except as resource-intensive add/drop devices~\cite{Brecht2015}. Applied temporal mode encodings may need multimode reshaping, for example, to match the central frequencies and bandwidths of PDC photons to the acceptance range of a solid-state memory interface~\cite{Kielpinski2011}, or to develop resource-efficient rotations and manipulations in the temporal mode basis. Initially, single-photon SFG was explored in the context of upconversion detectors, which efficiently shifts the frequency of photons from the telecom regime to the visible, where more efficient avalanche photodiodes exist~\cite{Vandevender2004,Thew2006}. While advances in superconducting nanowire detectors have eased telecom detection requirements, such processes have continued to find quantum applications, including frequency conversion for connecting quantum network nodes~\cite{Kielpinski2011,agha2014spectral,heinze2017photonic} and ultrafast signal gating~\cite{Kuzucu2008,Allgaier2017a,maclean2017direct}.  Multimode SFG processes have been shown to add little noise, evidenced through experiments which have confirmed entanglement preservation in time bin~\cite{Tanzilli2005} and polarisation~\cite{ramelow2012polarization,donohue2014ultrafast} degrees of freedom after frequency conversion and bandwidth manipulation.

For more general transformations, we can look to concepts from temporal imaging~\cite{Kolner1989,bennett1994temporal}, which describes manipulations to the temporal structure of light in much the same way that spatial imaging describes the actions of lenses and diffractive propagation. Temporal imaging systems require the ability to implement phase shaping in both the spectral and temporal domains. Spectral domain manipulations can be accomplished simply with phase-only pulse shaping or standard dispersion-compensation techniques~\cite{weiner2000femtosecond}, but temporal phase manipulation (often called ``time lensing'') is more difficult for sub-picosecond pulses, especially at the quantum level. Recently, groups have shown that dispersion and sum-frequency generation provides an effective toolbox for manipulating the bandwidth and time scale of PDC photons~\cite{lavoie2013spectral} as well as reshaping the time-frequency structure of entangled photon pairs~\cite{Donohue2016}. These techniques work in the exact opposite regime as the QPG, in that broad, non-restrictive phasematching is desired, i.e. all three fields must stay approximately group-velocity matched through the interaction. This often limits SFG-based time lenses to short nonlinear crystals, but the process can in principle reach high efficiency without the same time-ordering roadblocks as mode-selective measurement~\cite{Reddy2013,Donohue2015,Patera2015}. Note that temporal imaging can be accomplished in analogous ways through four-wave mixing~\cite{salem2008optical,Shi2017}. Alternatively, other groups have shown deterministic time lensing using electro-optic modulation~\cite{Karpinski2016,wright2017spectral,mittal2017} and cross-phase modulation~\cite{Matsuda2016}. Taking concepts from the work done on quantum temporal imaging and applying them to temporal-mode manipulation is an exciting direction for future research.

\section{Towards applications of temporal modes in quantum information science}\label{sec:five}

Finally, in this section, we outline experimental progress towards harnessing mode-selective upconversion for quantum technologies. The experiments referenced above have shown that quantum pulse gates can be realised with high efficiencies and high selectivities. In order to apply them for quantum signal processing, high signal-to-noise ratios are absolutely essential to separate quantum from classical signals and to protect resources such as entanglement and squeezing.

\begin{figure}
\centering
\includegraphics[width=1.\linewidth]{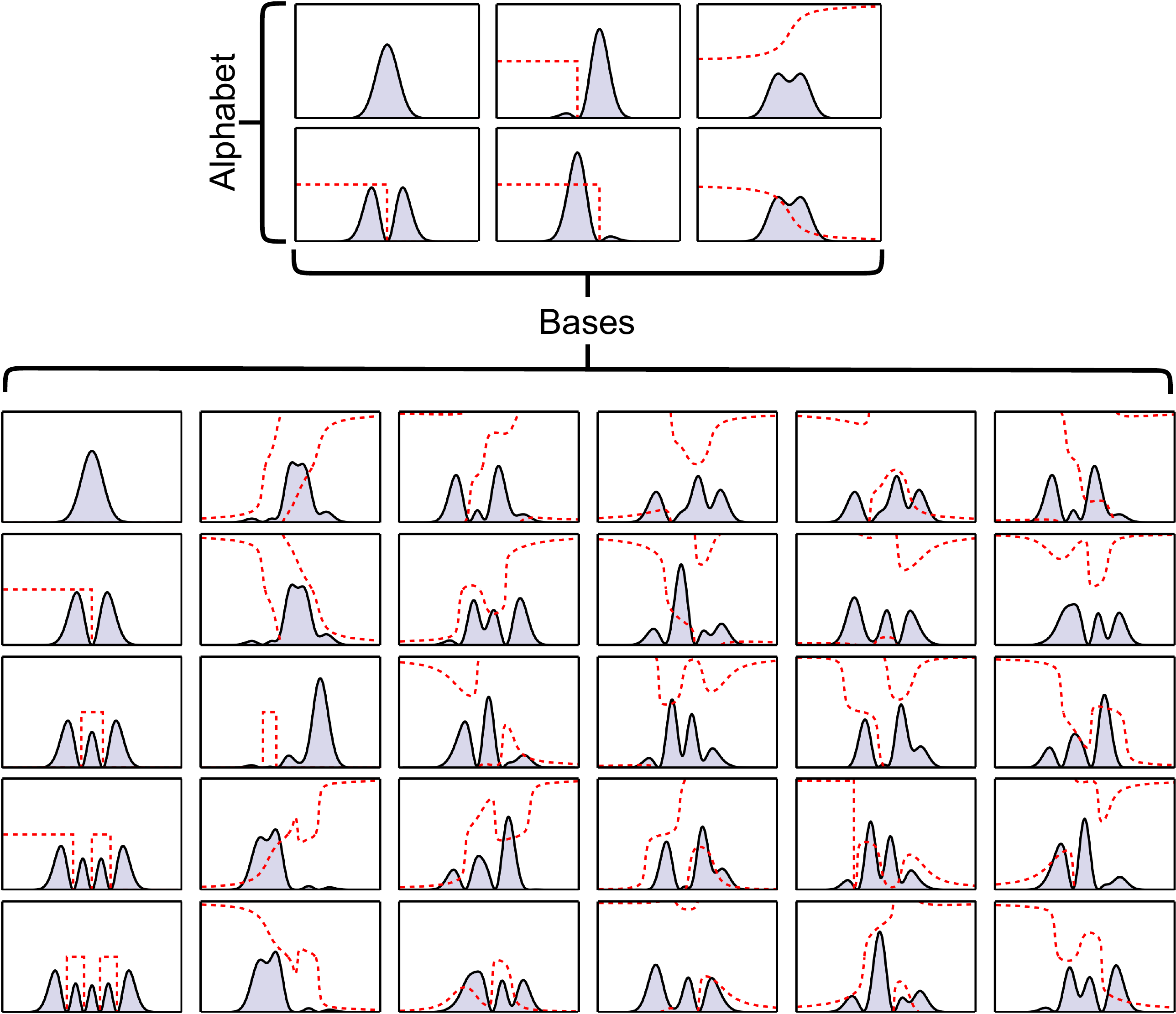}
\caption{Spectral field amplitudes spanning a complete set of mutually-unbiased bases for Hermite-Gauss modes in two (top) and five (bottom) dimensions~\cite{bandyopadhyay2002new,Ansari2017}. In order to completely access the Hilbert space, effective projections on all of these states must be realisable. The normalised spectral intensity is shown in grey and the red line corresponds to the spectral phase (on the interval 0 to $2\pi$).}
\label{fig:mubs}
\end{figure}

To exploit temporal modes as a high-dimensional coherent quantum resource, the selectivity must be maintained for a high-order alphabet as well as over the complete set of possible superposition states, as illustrated in Fig.~\ref{fig:mubs}. The security of quantum key distribution, for instance, relies entirely on the ability to measure complementary observables. For tomographic reconstruction of $d$-dimensional quantum states, projective measurements onto at least $d^2$ states spanning the total Hilbert space are required. A complete set of $d+1$ mutually unbiased bases~\cite{wootters1989optimal,bandyopadhyay2002new} provides a sufficient set of projections, examples of which in the Hermite-Gauss basis are shown in Fig.~\ref{fig:mubs}. High-dimensional two-qudit state tomography of entangled photons has been demonstrated with encodings in time-binned modes~\cite{nowierski2016tomographic,Ikuta2014}, spectral binned-modes~\cite{Schwarz2014,Kues2017}, and orbital angular momentum spatial modes~\cite{agnew2011tomography}. To avoid the intense resource devotion needed for full tomographic reconstruction, properties such as entanglement can be verified with witnesses instead~\cite{arrazola2012accessible,krenn2014generation,huang2016high,erker2017quantifying}. However, these techniques still require the ability to project in complementary bases.

Utilising the time-frequency degree of freedom for high-dimensional quantum information protocols has generally been confined to the context of time- or frequency-bin temporal modes, where the computational-basis modes are directly distinguishable in intensity. In particular, time bins have become the temporal-mode basis of choice behind the longest-distance Bell inequality violations over fibre networks~\cite{tittel1998violation,marcikic2004distribution,valivarthi2016quantum}, many commercial QKD systems~\cite{muller1997plug}, and high-dimensional entanglement-enabled quantum communication schemes~\cite{zhong2015photon,xie2015harnessing,ikuta2018four}. By passing a photonic signal through an unbalanced Mach-Zehnder interferometer, such that the reflected arm acquires an overall delay and adjustable phase relative to the transmitted, a superposition of arrival times can be prepared or measured~\cite{franson1989bell}. Extensions to higher dimensions have been realised with multi-path interferometers~\cite{thew2004bell}, cascaded Mach-Zehnder interferometers with different delays~\cite{Ikuta2014,ikuta2018four}, and time-to-polarisation conversion enabled by cross-phase modulation~\cite{nowierski2016tomographic}. However, the interferometers in the first two techniques require detectors with time resolution fine enough to separate non-interfering events, and the latter technique is limited in which superpositions can be directly measured. By using SFG with chirped inputs as a time-to-frequency converter, it has been demonstrated that projective measurements can be made on superpositions of time-bin photonic states on time scales well below detector resolution~\cite{donohue2013coherent}. While this technique was effective enough to convincingly violate a Bell inequality and reconstruct time-bin qubit density matrices, it is limited to a maximum efficiency of $1/d$ for a given projection.

SFG has also been key to frequency-bin encoded schemes, particularly those involving the recombination of a PDC photon pair in a second nonlinear crystal~\cite{ODonnell2009,Peer2005}. By creating spectrally entangled photons and slicing their spectra into bins, researchers have used this method to demonstrate novel high-dimensional encoding schemes~\cite{Lukens2014} and violate high-dimensional Bell inequalities~\cite{Schwarz2014}. However, since these experiments rely on recombination of the two photons, they are difficult to extend to quantum network applications. Recent work using low-noise electro-optic modulators to create sidebands from a frequency comb source has enabled projective measurements on frequency-bin entangled photons from frequency comb sources without needing the two photons to recombine~\cite{olislager2010frequency,jaramillo2017persistent,Kues2017}. These tools have been demonstrated to enable deterministic frequency-bin rotations~\cite{lukens2017frequency,lu2017electro} and fast feed-forward frequency shifting for spectrally multiplexed photon sources~\cite{puigibert2017heralded}.

The dispersion-engineered techniques outlined in Sec.~\ref{sec:three} have the key advantage that, so long as the transfer function of \eqref{eq:transfunc} remains separable, they are capable of projecting onto temporal modes in arbitrary bases, including both the binned modes and field-overlapping pulse modes. To be effective for high-dimensional quantum protocols, dispersion-engineered mode-selective SFG must be both low-noise and coherent, in the sense that it remains effective for not only the basis modes but also general superpositions. Progress has been made towards applying the quantum pulse gate to photonic state characterisation and manipulation, but it remains an active field of research. 

Using the configuration of Ref.~\cite{Brecht2014} with input from a spectrally pure PDC source, it was confirmed that the QPG output maintains nonclassical photon number correlations (i.e. the heralded $g^{(2)}$ of both the input and register modes was measured to be $0.32\pm0.01<1$)~\cite{Allgaier2017}. By shaping the QPG pump over a tomographically complete set of TMs, this setup has been used to reconstruct the one-qudit TM density matrix of PDC photons varied from single- to multimode configurations, with both intensity- and phase-correlated multimode structure~\cite{Ansari2016}. However, worse performance was noted for higher-dimensional reconstructions. The device's performance was fully characterised through temporal-mode detector tomography~\cite{Ansari2017}, which showed that a system based on a 17-mm PPLN waveguide could reconstruct the TM density matrix in seven dimensions with a fidelity higher than 80\%. By calibrating the QPG with this detector tomography, the reconstruction algorithm could be altered to reconstruct randomly generated seven-dimensional coherent superpositions of temporal modes with a fidelity of $(98.8\pm0.4)\%$. These experiments are to-date the only dispersion-engineered TM measurements performed with a quantum light source rather than attenuated coherent light.

In a continuous-variable context, where quantum information is encoded in field quadratures rather than superpositions of discrete qudit states, temporal modes still serve an important purpose in SPOPOs. However, for these to work, continuous-variable operations must operate in a mode-selective fashion. The group of N.~Treps showed that QPG techniques can work as a mode-selective photon subtractor, a key non-Gaussian component of the continuous-variable toolkit~\cite{Averchenko2014,Averchenko2016,Ra2017}. Since the SPOPO emits squeezed light over many temporal modes, a mode-selective beam splitter is necessary to ensure that the heralded photon subtraction is matched to the desired temporal mode. Using a noncollinear frequency-degenerate phasematching in bulk bismuth borate (BiBO) supplemented with spectral filtering and shaped weak coherent states ($\bar{n}<1$), Ra et al. were able to reconstruct the temporal-mode subtraction matrix in both the spectral bin and Hermite-Gauss basis~\cite{Ra2017}, which characterises the modal purity of the subtraction process. For a seven-dimensional HG superposition, the subtraction matrix was found to have a purity of 96\% regardless of whether the signal was bright or on the single-photon level. Since the photon-subtraction method requires weak coupling in order to minimally disturb the quantum state, a QPG with a low efficiency (0.1\%) was used, equivalent to a low-reflectivity beam splitter ~\cite{Averchenko2014}.

\section{Outlook and Challenges}\label{sec:conc}
We have shown that dispersion-engineered waveguides provide a capable toolbox for generating and measuring photon temporal modes. By constructing photon-pair sources simultaneously pure in both spatial and temporal degrees of freedom as shown in Section~\ref{sec:two}, it is possible to efficiently create pure heralded single photons, capable of providing the high-visibility quantum interference necessary for multiphoton quantum logic. By exploiting the group-velocity matching of these systems, it was also shown that the temporal shape and entangled structure of the temporal modes can be customised, providing a versatile resource for quantum state engineering. In Section~\ref{sec:three}, it was shown that these same engineered techniques can be applied to sum-frequency generation, providing the necessary tools to manipulate and measure this structure. In Sections~\ref{sec:four}~and~\ref{sec:five}, we outlined the considerable experimental progress that has been made towards realising this toolbox.

Many challenges remain to push toward practical application. Temporal-mode selective devices have been demonstrated in the sub-picosecond or few-picosecond regime, where commercially available pulse shapers exist. Such time scales are natural for PDC processes, but come with difficult synchronisation challenges for long-distance quantum communication or entanglement distribution. Moving to longer, less jitter-sensitive regimes through memory-based interfaces or resonant cavities~\cite{Reddy2017} relaxes this concern, but increases the burden of pulse shaping. Four-wave mixing techniques have more complicated noise landscapes for quantum tasks, but offer considerably longer interaction lengths and are currently under-studied for temporal-mode management. In all cases, for high-dimensional tasks, devices which isolate a single temporal mode are difficult to scale, requiring multiple shaped pulses and physical media to construct a multi-output measure. Techniques which demultiplex a set of pulsed temporal modes into spatial or spectral bins, equivalent to the orbital angular momentum mode sorter in space~\cite{berkhout2010efficient}, are essential to scale these techniques to high-dimensional networks. A promising avenue for these temporal-mode demultiplexers is through multi-peak phasematching structures~\cite{asobe2003multiple,silver2017spectrally}.

By accessing the temporal mode structure of quantum light, we can open a new frontier in photonic quantum information science. By tailoring PDC sources to directly generate pure photon pairs, an important step towards scalable quantum networks has been taken. With measurements sensitive to the time-frequency structure in arbitrary phase-dependent bases, quantum pulse gates may open the door to novel ultrafast measurement schemes. We have outlined some of the significant advances that have been made in the past ten years from numerous researchers across the globe. With an active and engaged community, we eagerly anticipate the next ten.

\section{Funding}
This work was supported by the Deutsche Forschungsgemeinschaft (DFG) Sonderforschungsbereich TRR 142, the Natural Sciences and Engineering Resource Council of Canada (NSERC), and from the European Union's (EU) Horizon 2020 research and innovation program under Grant Agreement No. 665148.

\section{Acknowledgments}
The authors thanks M. Santandrea, D.V. Reddy, H. Herrmann, and E. Meyer-Scott for useful discussions and suggestions.



%

\end{document}